\documentclass[journal, comsoc]{IEEEtran}
\IEEEoverridecommandlockouts

\usepackage{hyperref}
\hypersetup{
     colorlinks = true,
     linkcolor = blue,
     anchorcolor = red,
     citecolor = magenta,
     filecolor = blue,
     urlcolor = black,
}

\usepackage{cite}
\usepackage{amsmath,amssymb,amsfonts}
\usepackage{algorithmic}
\usepackage{graphicx}
\usepackage{textcomp}
\usepackage{xcolor}
\usepackage{authblk}
\usepackage{amsmath,amssymb}
\usepackage{url}
\usepackage{booktabs,makecell,tabularx}
\usepackage{threeparttable}
\usepackage{graphicx}

\usepackage{soul}

\usepackage{multirow}
\usepackage{subfigure}
\usepackage[linesnumbered,ruled,vlined]{algorithm2e}
\def\BibTeX{{\rm B\kern-.05em{\sc i\kern-.025em b}\kern-.08em
    T\kern-.1667em\lower.7ex\hbox{E}\kern-.125emX}}

\begin{document}

\title{ByCAN: Reverse Engineering Controller Area Network (CAN) Messages from Bit to Byte Level}

\author{Xiaojie~Lin,
        Baihe~Ma,
        Xu~Wang,~\IEEEmembership{Member,~IEEE},
        Guangsheng~Yu,
        Ying~He,~\IEEEmembership{Senior Member,~IEEE},
        Ren~Ping~Liu,~\IEEEmembership{Senior Member,~IEEE},
        and~Wei~Ni,~\IEEEmembership{Fellow,~IEEE}
    \IEEEcompsocitemizethanks{
        \IEEEcompsocthanksitem X. Lin, B. Ma, X. Wang, Y. He, and R. P. Liu are with the Global Big Data Technologies Centre, University of Technology Sydney, Australia. \protect
        E-mail: \{xiaojie.lin, baihe.ma, xu.wang, ying.he, renping.liu\}@uts.edu.au
        \IEEEcompsocthanksitem G. Yu and W. Ni are with Data61 CSIRO, Sydney, Australia. \protect
        E-mail: \{saber.yu, wei.ni\}@data61.csiro.au
    }
}

\maketitle

\begin{abstract}
As the primary standard protocol for modern cars, the Controller Area Network (CAN) is a critical research target for automotive cybersecurity threats and autonomous applications.
As the decoding specification of CAN is a proprietary black-box maintained by Original Equipment Manufacturers (OEMs), conducting related research and industry developments can be challenging without a comprehensive understanding of the meaning of CAN messages.
In this paper, we propose a fully automated reverse-engineering system, named ByCAN, to reverse engineer CAN messages.
ByCAN outperforms existing research by introducing byte-level clusters and integrating multiple features at both byte and bit levels.
ByCAN employs the clustering and template matching algorithms to automatically decode the specifications of CAN frames without the need for prior knowledge.
Experimental results demonstrate that ByCAN achieves high accuracy in slicing and labeling performance, i.e., the identification of CAN signal boundaries and labels.
In the experiments, ByCAN achieves slicing accuracy of $80.21\%$, slicing coverage of $95.21\%$, and labeling accuracy of $68.72\%$ for general labels when analyzing the real-world CAN frames.
\end{abstract}


\section{Introduction}
\label{intro}

{\color{black}
Massive volumes of data are now generated and transmitted via In-Vehicle Networks (IVNs) as modern cars are equipped with more in-vehicle Electronic Control Units (ECUs) with communication capabilities~\cite{leen2002expanding, wu2019survey}. 
The Controller Area Network (CAN) protocol was firstly developed by Bosch in the 1980s~\cite{leen2002expanding} and serves as the de facto standard protocol for connecting ECUs embedded in cars~\cite{MK,Zeng,avatefipour2018state}. 
The standard structure of the CAN frame is composed of the start of frame, arbitration field, control field, data field, CRC field, ACK field and end of frame, as shown in Fig.~\ref{CANframe}.
While the CAN protocol has a standardized frame structure, understanding the protocol's utilization for signal transmission remains challenging.
This is because Original Equipment Manufacturers (OEMs) encode the signals within the CAN frames' data fields (data payloads) in proprietary ways that vary among OEMs, vehicle models, and years~\cite{buscemi2023survey}.

\begin{figure}[t]
        \centering
        \includegraphics[width=0.95\columnwidth]{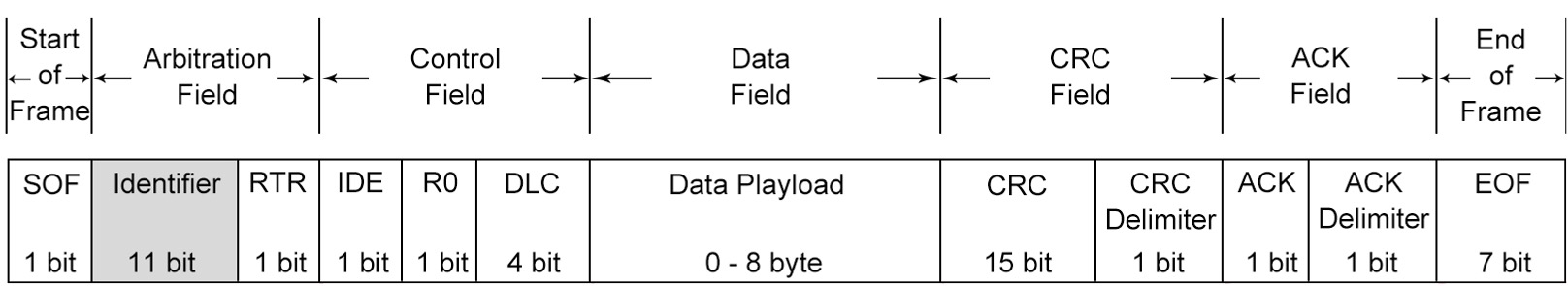}
        \caption{{\color{black}Standard CAN frame format: 1-bit Start of Frame (SOF), 11-bit ID, 1-bit Remote Transmission Request (RTR), 1-bit Identifier Extension Bit (IDE), 1-bit Reverse R0, 4-bit Data Length Code (DLC), 0 to 8 byte data payload, 15-bit Cyclic Redundancy Check (CRC), 1-bit CRC delimiter, 1-bit Acknowledge (ACK), 1-bit ACK delimiter, and 7-bit End of Frame (EOF). }}
        \label{CANframe}
\end{figure}

Understanding and decoding the data payloads of CAN frames is the first step to extracting the essential information to develop autonomous applications or explore automotive cybersecurity threats, such as vehicle location detection~\cite{ma2022personalized, ma2022new, ma2023vehicle, ma2023location}.
Many studies have been dedicated to developing an automated Reverse Engineering (RE) system for CAN with minimum human interactions to uncover the communication patterns of CAN signals and deduce the decoding specification of the data payloads of CAN frames~\cite{nolan2018unsupervised, frassinelli2020know, verma2018actt, Markovitz, Marchetti, Pese, buscemi2021canmatch, Verma, choi2021enhanced}.
This could allow penetration testing and digital forensics on the CAN bus.
Automotive insurance companies benefit from RE systems because the result of those systems provides a way to evaluate cybersecurity risks, as well as to discover driver fingerprinting for usage-based insurance~\cite{buscemi2023survey}. 
}

{\color{black}
Existing methodologies leverage CAN frames and additional data sources, such as the Global Positioning System (GPS)~\cite{frassinelli2020know}, Inertial Measurement Unit (IMU)~\cite{Pese}, DBC files~\cite{buscemi2021canmatch}, and On-Board Diagnostics II (OBD-II) data~\cite{blaauwendraad2020automated}, to decode the specification of CAN signals.
Among these, the OBD-II diagnostic data is easy to access via the OBD-II port, as all modern cars are equipped with the OBD-II diagnostic system.
OBD-II diagnostic data can be converted into human-readable accurate vehicle data with public formulas to be used in the matching process for associating semantic meanings with CAN signals. 
Both OBD-II diagnostic data and regular CAN frames can be collected from the OBD-II port.
The RE systems can leverage both CAN and OBD-II diagnostic data to create a comprehensive dataset for reverse engineering purposes, eliminating the need for additional measurement equipment like IMUs.
}

\begin{figure}[t]
        \centering
        \includegraphics[width=0.85\columnwidth]{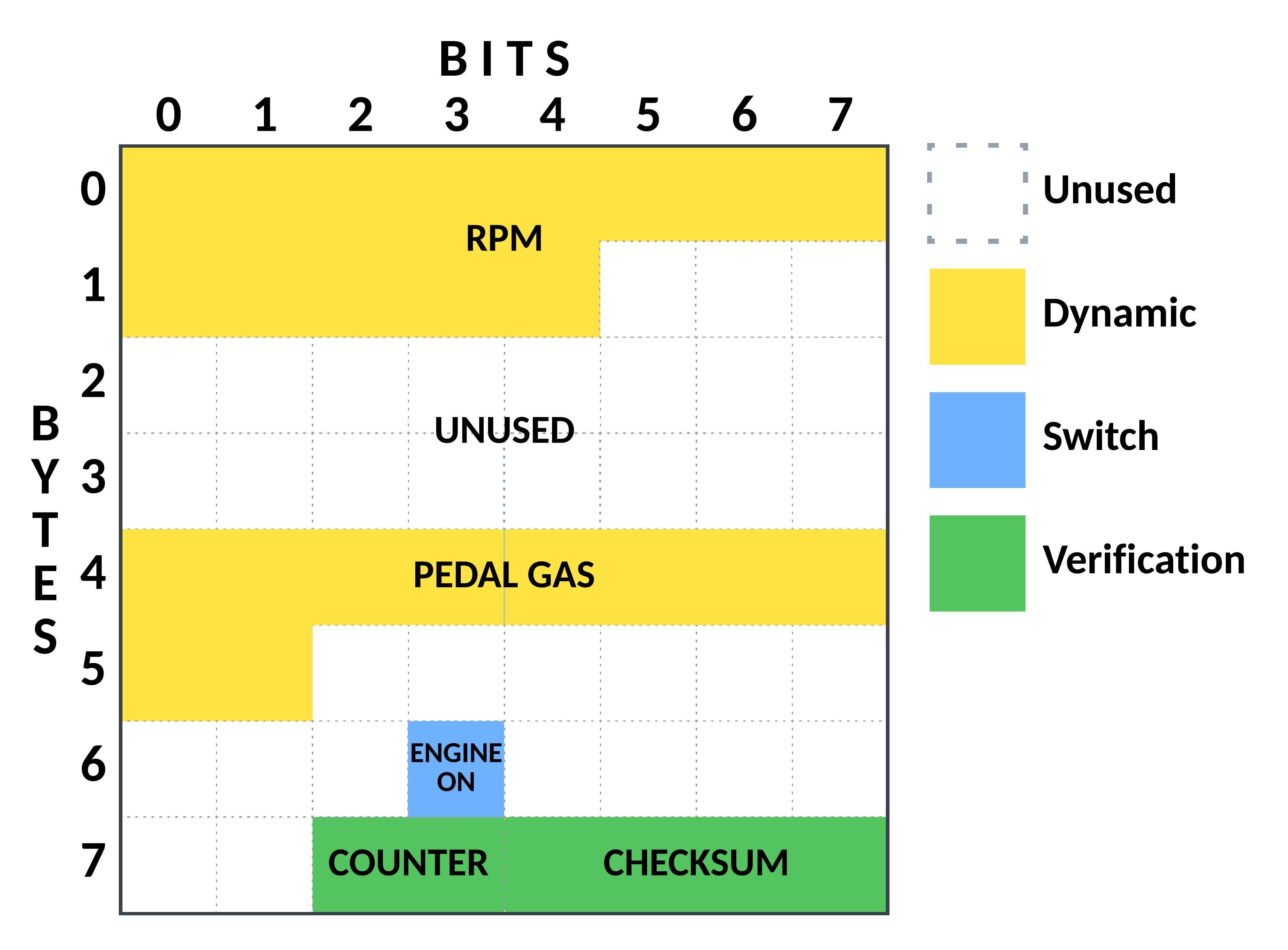}
        \caption{Sample DBC: Mazda 3, Year 2019, CAN ID of 0x01A~\cite{openDBC}. CAN signals may occupy more than one byte, e.g., engine speed and pedal gas, which can be found by the yellow $Dynamic$ CAN signals that take more than 8 bits consecutively. 
        CAN signals often align with whole-byte offsets, such as the \textit{Dynamic} signals aligned to the left and the \textit{Verification} signals aligned to the right.}
        \label{slicing:showcase}
\end{figure}

The primary objective of a CAN RE system is to identify the boundaries of CAN signals within a CAN frame payload field and to determine the associated signal labels.
CAN messages are structured into frames, and the CAN frames of different CAN IDs have different lengths of the data payload.
The payload servers as a container for multiple CAN signals, each with distinct bit positions and lengths, as illustrated in Fig.~\ref{slicing:showcase}.
Before labeling CAN signals, the CAN signal slicing process first identifies the CAN signals' boundaries.
The bit-flip rate, representing the frequency at which bits change from 1 to 0 or vice versa, is the primary feature in existing methodologies for slicing CAN signals~\cite{buscemi2023survey}.
However, CAN signal slicing with only the bit-flip rate cannot precisely differentiate varying CAN signals.
For example, distinguishing between \textit{Dynamic} (e.g., engine speed signals)  and \textit{Verification} signals (i.e., \textit{Checksum} and \textit{Counter}) is challenging because their bit-flip rates are similar.
\textcolor{black}{By relying solely on bit-level features~\cite{verma2018actt, Markovitz, Marchetti, Pese, buscemi2021canmatch, Verma, choi2021enhanced}, existing RE systems can result in excessive slicing of CAN signals.}

{\color{black}
The byte-level patterns of CAN signals are considered in this paper based on the observation of DBC files from the OpenDBC repository~\cite{openDBC}.
The DBC files from the OpenDBC repository cover $82$ cars, ranging from 2001 to 2023 for $19$ distinct OEMs, such as Honda, Mazda, Mercedes Benz, and Toyota.
There are $34,495$ CAN signals covered in the DBC files from the OpenDBC, within which $98.25\%$ of CAN signals are shorter than or equal to $16$ bits of signal length.
}
In summary, we found the patterns of CAN signals as follows:

\begin{itemize}
    \item Pattern 1: CAN signals usually occupy consecutive bits within a CAN frame, necessitating the signal slicing to locate CAN signals.
    \item Pattern 2: CAN signals may occupy more than one byte, highlighting the need for identifying signal boundaries at the byte level.
    \item Pattern 3: CAN signals often align with whole-byte offsets, suggesting the advantage of initial byte level slicing.
\end{itemize}
These insights are indicated in Fig.~\ref{slicing:showcase}.
Thereby, the byte-level clustering step can help in finding the boundaries of CAN signals at byte level before slicing CAN signals at bit level to enhance accuracy. 

This paper introduces a fully automated RE system named ByCAN to decode the data payloads of CAN frames using only the collected CAN frames and OBD-II diagnostic data with clustering-based matching learning and template matching algorithms.
Unlike previous studies, ByCAN processes CAN signals at both byte and bit levels to reduce inaccurate in signal slicing, particularly, to avoid the issue of excessively sliced signal boundaries. 
To better capture the identified three patterns, ByCAN proposes multiple features, such as flip rate, average value, and distinct value ratio, at both the byte and bit levels. 
The system's performance is assessed on CAN messages captured from real-world vehicles.
The key contributions of the proposed ByCAN are summarized as follows:
\begin{itemize}

    \item We design a fully automated RE system using easily accessible data sources, i.e., CAN frames and OBD-II diagnostic data, to decode the specifications of CAN signals.

     \item We propose new features of CAN signals at both byte and bit-block levels to capture high-level patterns of CAN signals.
     Compared to the restrictive single-bit features, the byte and bit-block level features can capture the dynamics across a broader range of values, revealing complex and subtle patterns in CAN signals. To the best of our knowledge, this is the first attempt to use byte-level clusters and features to slice CAN signals.
    
    \item We propose a new slicing algorithm featuring Density-Based Spatial Clustering of Applications with Noise (DBSCAN), which eliminates the need for an estimated count of CAN signals. We also introduce a new template matching process utilizing Dynamic Time Warping (DTW), specifically designed to align large-volume CAN signals with the significantly less frequent OBD-II diagnostic data.

\end{itemize}

The rest of the paper is organized as follows.
Section~\ref{backgrounds} introduces relevant background information, and Section~\ref{related} gives the related works. Section~\ref{model} presents the proposed system. Section~\ref{performance} evaluates the proposed system, and Section~\ref{conclusion} concludes the paper.

{\color{black}
\section{Background}
\label{backgrounds}
\subsection{Controller Area Network}\label{CAN}

A CAN frame contains the start of frame, arbitration field, control field, data field, CRC field, ACK field and end of frame, as in Fig.~\ref{CANframe}.
ECUs have their own CAN IDs as unique identifiers~\cite{leen2002expanding}. 
The CAN ID of each frame is the reference to arbitrate the priority of concurrent CAN frames. 
Real-time information of vehicle states (e.g., vehicle speed, indicator status and checksum) is enclosed in the data payload. 
The length of the data payload is up to 8 bytes. One CAN frame may contain multiple CAN signals of different functions in the data payload. For instance, different CAN signals related to doors and trunks (e.g., front left door open, rear right door open, and trunk open) can be located in the same CAN frame. 
OEMs have different decoding specifications of CAN frames such that different car models may use different CAN IDs for the same CAN signals.

\subsection{OBD-II Diagnostic Data}
OBD-II is an automotive diagnostic system developed in 1992 and installed mandatory for all cars built after 1996 in the US~\cite{hermawan2020acquisition}.
The primary function of OBD-II diagnostic messages is to monitor the vehicle health state and enable the vehicle to communicate its operational information through standardized Diagnostic Trouble Codes (DTCs)~\cite{mccord2011automotive}.
OBD-II diagnostic data use Parameter IDs (PIDs) to specify different vehicle states be requested or transmitted. 
These PIDs govern a range of data from vehicle speed to sensor data for fuel system status with a published standard to interpret related signal messages.
The same vehicle state, e.g., vehicle speed, can be obtained from the regular CAN frames and the requested OBD-II diagnostic data but following the CAN encoding and OBD-II diagnostic encoding, respectively.

\subsection{Collecting CAN Frames and OBD-II Diagnostic Data}

CAN frames and OBD-II diagnostic data are accessed via the OBD-II port that typically located under the dashboard. 
By connecting CAN analyzers to the OBD-II port, CAN frames can be recorded passively, and meanwhile OBD-II diagnostic data can be requested actively.
The CAN frames from the OBD-II port follow the same format as those over the in-vehicle CAN bus.
This is because the OBD-II port is hard-wired to the CAN bus and provides CAN high and CAN low for passive listening. 
There could be a CAN gateway between CAN bus and OBD-II port which regulates the transfer of CAN data and can filter out CAN frames based on specific CAN IDs before sending through the OBD-II port~\cite{zhang2016controlling, ammar2020securing}.
Different methods can be used to gain direct access to the CAN bus, e.g., via the instrument pack or the rear of the radio, to bypass the CAN gateway and obtain all CAN signals.
The only difference between the CAN frames collected from the OBD-II port and those from the CAN bus wires is the number of CAN IDs and CAN signals in the collected dataset.
As existing RE systems focus on analyzing CAN IDs independently, the absence of specific CAN IDs does not impede the RE process on other accessible CAN IDs.
Since the OBD-II port is the common access point without damaging cars, most existing works promise utility and applicability by collecting CAN frames via the OBD-II port.

\subsection{CAN Database Container}\label{DBC}

The CAN Database Container (DBC) is a widely adopted format developed for the storage and interchange of the decoding specifications of CAN frames' data payload.
DBC files are in the ASCII format to keep it human-readable and easy to manipulate programmatically.
The DBC file describes the CAN IDs, the lengths of the data payloads of CAN frames, and the names of CAN signals.
The DBC file also encapsulates the signal details, including the names, bit start position, length, endianness, scale, offset, range and unit of CAN signals associated with CAN ID.
However, DBC files vary across different vehicle models and makers~\cite{young2020towards}. 
OEMs rarely release DBC files due to security concerns, which makes the decoding specification of the data payloads of CAN frames a black box~\cite{rajapaksha2023ai}.

\subsection{DBSCAN Clustering Algorithm}

Density-Based Spatial Clustering of Applications with Noise (DBSCAN) clustering algorithm was first introduced in 1996 and widely used to recognize data patterns~\cite{ester1996density}.
DBSCAN identifies the core points with many nearby neighbors, the border points that are reachable from core points but not dense themselves, and the outliers that lie alone in low-density regions~\cite{schubert2017dbscan}.
Unlike the K-Means, DBSCAN does not require the specification of the number of clusters as an input parameter.
Instead, the DBSCAN clustering algorithm only requires two parameters: The search radius $\epsilon$ around each point to find neighbors and the minimum number of points \textit{minPts} to determine a dense region for the core point.
The number of clusters is determined by $\epsilon$, \textit{minPts} and the data itself, which makes DBSCAN particularly useful for the clustering tasks with unknown data structures.
}

\begin{table*}[htb]
  \centering
  \caption{{\color{black}Comparison of reverse engineering systems}}
  \label{tab:models_compare}
  \begin{threeparttable}

  \begin{tabular}{p{0.11\linewidth}p{0.15\linewidth}p{0.25\linewidth}p{0.2\linewidth}p{0.17\linewidth}}
    \toprule
      & Approaches & \multicolumn{2}{c}{Input Features of CAN Message} & Data Source \\
      \cmidrule(lr){3-4}
      & & Bit-level & Byte-level & \\

    \midrule
    
    READ ~\cite{Marchetti} & Taxonomy & Bit-flip Rate & -  & CAN\\
    \addlinespace[5pt]
    
    LibreCAN ~\cite{Pese} & Taxonomy & Bit-flip Rate & - & CAN, GPS, IMU, OBD-II\\
    \addlinespace[5pt]
  
    CANMatch~\cite{buscemi2021canmatch} & Unsupervised Learning, & Bit-flip Rate & - & CAN, DBC Files\\
    & Frame Matching$^1$ & & & \\
    \addlinespace[5pt]

    CAN-D ~\cite{Verma} & Supervised Learning, & Bit-flip Rate, & - & CAN, OBD-II\\
    & Unsupervised Learning & Neighbouring Bit's Values and Flips$^2$ & & \\
    \addlinespace[5pt]

    \textbf{Our work} & Unsupervised Learning,& Bit-flip Rate, & Byte Flip Rate, & CAN, OBD-II \\
    & Template Matching & Bit-block Flip Rate, & Average Byte Value, & \\
    &  & Average Bit/Bit-block Value, & Distinct Byte Value Ratio & \\
    & & Distinct Bit-block Value Ratio & & \\

    \bottomrule
  \end{tabular}
  \begin{tablenotes}
  \footnotesize
  \item[1] CAN signals whose ID can be matched in DBC files (ground truth) are directly reverse-engineered by using the ground truth results, which makes CANMatch largely depend on the ground truth.

  \item[2] The neighboring bit's values and flip rates are not directly used but encoded.
  
  \end{tablenotes}
  \end{threeparttable}
\end{table*}

\section{Related Work}
\label{related}
CAN RE systems deduce CAN signals' boundaries and labels by slicing and labeling CAN signals from collected CAN frames~\cite{buscemi2023survey}.
Existing works have developed from the manual systems~\cite{checkoway2011comprehensive, woo2014practical, buscemi2020data}, which require significant human effort, to automated systems that analyze CAN signal patterns from collected CAN frames for more efficient slicing and labeling~\cite{Markovitz, choi2021enhanced, Marchetti, Pese, Kang, Hoog, Salvador, Wen, Verma, buscemi2021canmatch}.
Markovitz and Wool are the first to develop an automated RE system and evaluate the performance with the simulated CAN frames~\cite{Markovitz}. 
Subsequently, Marchetti and Stabili develop the READ system and evaluate the performance with real-world cars~\cite{Marchetti}.
Machine learning algorithms are used to enhance the automated RE when analyzing the real-world traffic trace of CAN frames where most CAN signals are transmitted at millisecond frequencies~\cite{bi2022bit}.

The bit-level features, such as the bit-flip rates of CAN signals, are largely utilized in the existing automated CAN RE system. 
The RE system named READ calculates the bit-flip rate for each bit in the CAN frame's data payload and builds up the magnitude array to slice CAN signals~\cite{Marchetti}.
READ identifies CAN signals' boundaries by searching the magnitude array for pairs of consecutive bits where the first exhibits a higher bit-flip magnitude than the second, marking such instances as potential signal boundaries.
Based on READ, Pesé et al. use the bit-flip rate and magnitude array to determine CAN signals' boundaries but introduce a specific percentage threshold for the bit-flip rate decrease to refine the pair identification~\cite{Pese}.
The CANMatch system focuses on the bit-flip rate only but extends the use of bit-flip rate by incorporating the endianness of CAN signals at the byte level~\cite{buscemi2021canmatch}. In~\cite{Verma}, the CAN-D system is developed to slice and label CAN signals with the input of multiple bit-level features, e.g., the bit-flip rate, two-bit distributions, and entropy.
In~\cite{young2020towards}, the value of each byte in the CAN frame of different CAN IDs is used to find the clusters of various vehicle functions.

Existing RE systems utilize byte values or endianness merely as supplementary to bit-level slicing, thus neglecting additional byte-level patterns of CAN signals. 
The byte-level features indicate different patterns of various CAN signals~\cite{liang2022leveraging, lin2022multi}. 
According to our observation, CAN signals typically occupy up to two bytes, and those CAN signals associated with similar vehicle functions often consecutively reside within one or two bytes. 
By examining the byte-level patterns, such as the flip rate and average value of the byte, CAN signals can be initially separated at the byte level to prevent excessive slicing.

{\color{black}
The existing works and the proposed ByCAN system are compared in Table~\ref{tab:models_compare}.
The RE systems have developed from the basic taxonomy approaches to machine learning-based methods.
We apply unsupervised learning and template matching in the proposed ByCAN system for clustering and labeling steps, respectively.
In prior works, the bit flip rate~\cite{Marchetti, Pese,buscemi2021canmatch,Verma} or the set of encoded features of neighbouring bit's values and flips~\cite{Verma} are used to slice and label CAN signals.
In contrast to existing methods, which have typically overlooked the byte-level patterns of CAN signals, ByCAN highlights the byte-level patterns and applies unsupervised learning algorithms in both slicing and labeling procedures.
ByCAN introduces multiple byte-level and bit-level features for byte-level signal clustering and bit-level signal slicing. 
Additional data sources, such as GPS~\cite{Pese}, DBC file~\cite{buscemi2021canmatch}, or OBD-II diagnostic data~\cite{Verma}, are examined in existing works to enhance the system performance.
The proposed system only requires CAN frames and the request for OBD-II diagnostic data that can be commonly collected from cars directly.
No prior knowledge, such as the DBC file used in CANMatch~\cite{buscemi2021canmatch} is required in ByCAN to slice and label CAN signals, which ensures the applicability of the proposed system for unknown cars and new models of cars.
The slicing and labeling performance of ByCAN is evaluated with CAN frames collected from real-world cars.
}

\begin{figure*}[t]
        \centering
        \includegraphics[width=1.6\columnwidth]{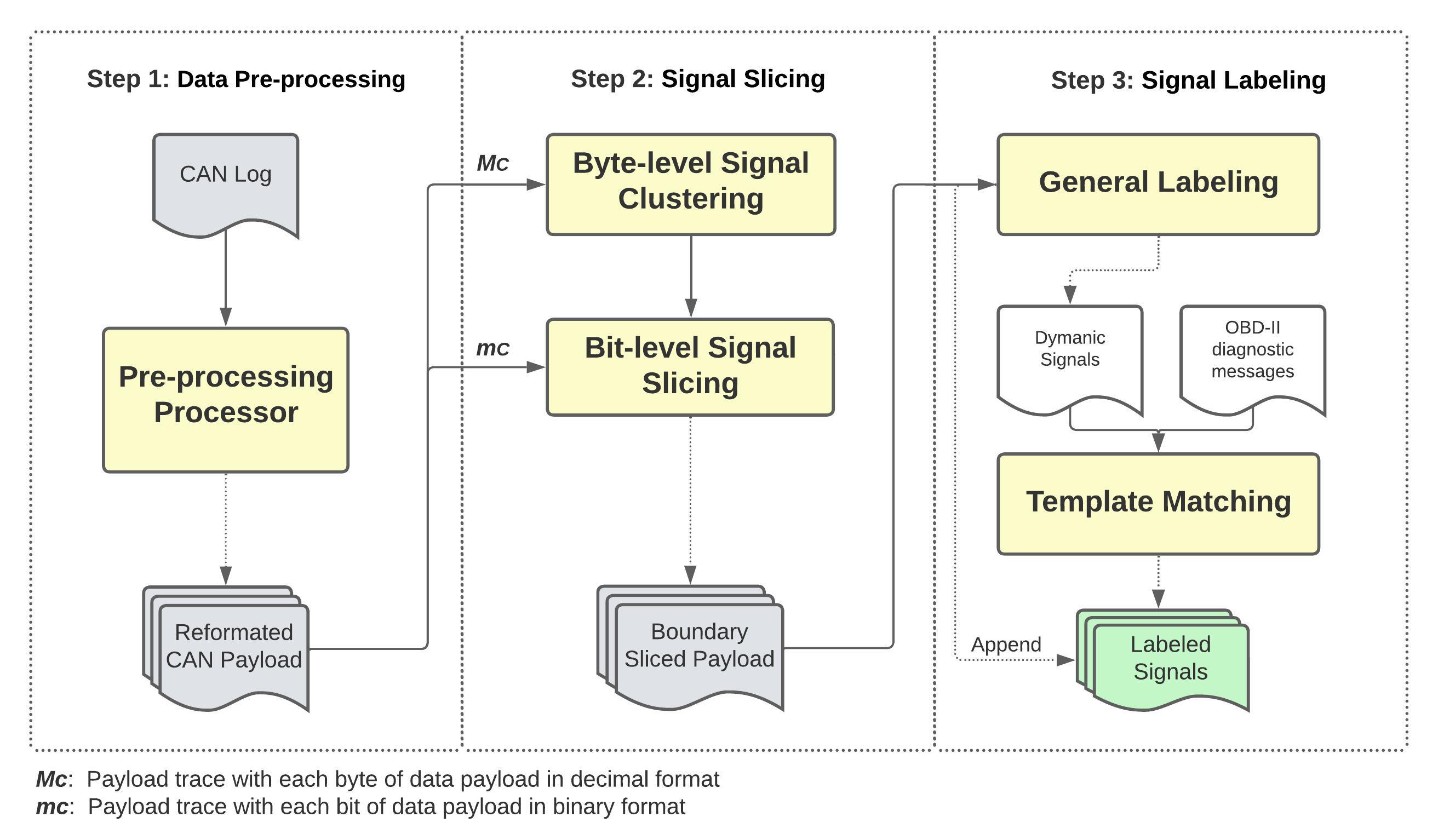}
        \caption{ByCAN system: \textcolor{black}{In the data pre-processing procedure, CAN messages are grouped by CAN ID first. Then, the grouped CAN messages are reformatted into the trace $M_C$ and $m_C$ with CAN frames' data payloads segmented at byte level and bit level, respectively.} 
        In the signal-slicing procedure, the byte-level CAN signal features are extracted to deduce byte-level signal clusters.
        The bit-level CAN signal boundaries are further sliced within each byte-level cluster using proposed signal features at the bit level.
        In the signal labeling procedure, the sliced CAN signals are first labeled as general categories (i.e., \textit{Unused}, \textit{Switch}, \textit{Dynamic} and \textit{Verification}).
        Finally, the descriptive labels are identified by applying the template matching algorithm to measure the similarity between the Dynamic signals and OBD-II diagnostic messages.}
        \label{model_flow}
\end{figure*}

\section{Proposed System}
\label{model}

The proposed CAN RE system, ByCAN, is demonstrated in Fig.~\ref{model_flow}. 
To the best of our knowledge, ByCAN is the first to utilize multiple features at both the byte and bit levels to identify CAN signals' boundaries and labels. 
The introduced features capture the characteristics of CAN signals and enhance the performance of the proposed ByCAN.
ByCAN consists of three procedures: Data Pre-processing, Signal Slicing, and Signal Labeling. 

{\color{black}
\subsection{Features Identification}
Based on empirical observations of the ground-truth DBC files and the collected CAN traces, the proposed system slices and labels CAN signals as \textit{Unused}, \textit{Switch}, \textit{Dynamic} and  \textit{Verification} types.
The \textit{Unused} CAN signals stand for the bits that are not taken or the values of the CAN signals are not changed during the communication period.
The \textit{Switch} CAN signals stand for the CAN signals that represent vehicle states whose values change rarely during the whole communication period.
The \textit{Dynamic} CAN signals stand for the vehicle's kinematic states, such as vehicle speed, which are quite active signals.
The \textit{Verification} CAN signals stand for the CAN signal counters or checksum that keep changing for every transmitted CAN frame.

Unlike previous works that primarily focus on the bit-flip rate or other bit-level features, our system comprehensively analyzes both byte and bit characteristics. 
According to the characteristics of different CAN signals, we propose new features to distinctly categorize CAN signals, including the flip rate, the average value, and the distinct value ratio for each byte, bit, and bit-block, as summarized below.

\subsubsection{Flip Rate}
The flip rate quantifies the variability of CAN signals across a sequence of CAN frames, which is fundamental for detecting the clusters of CAN signals at the byte or bit level within the data payloads of the CAN frames. 
Clusters with \textit{Dynamic} or \textit{Verification} CAN signals have high bit-flip rates, bit-block flip rates, and byte flip rates.
\textit{Dynamic} CAN signals typically represent continuous variables, either incrementing or decrementing by the base unit, whereas \textit{Verification} CAN signals pertain to discrete variables. 
This is because \textit{Dynamic} CAN signals capture vehicle kinematic states, which consistently change while driving. 
Consequently, the \textit{Verification} CAN signals are updated with each CAN frame.

\subsubsection{Average Value}
The use of the average value as a clustering parameter for CAN signals is substantiated by the observation that different types of CAN signals exhibit distinct patterns in their average values.
For example, the average value of vehicle speed is usually between 0 and 100 km/h, while the average value of engine speed usually ranges between 600 and 5000 RPM.
Therefore, the average value of the CAN signals could reveal the semantic meaning of the CAN signals.

\subsubsection{Distinct Value Ratio}
By measuring the number of values that are utilized actively, the distinct value ratios indicate the operational characteristics of the CAN signals, such as frequency of change and data density. 
The distinct value ratios indicate the functional complexity and utilization of the byte or the bit-block, revealing whether the byte or bit-block is predominantly occupied by dynamic values, static values, or a combination thereof. 
\textit{Dynamic} CAN signals may vary within a range constrained by vehicle kinematic states, such as vehicle speed, while verification signals can assume all possible values, as seen with counter signals.
Thus, \textit{Verification} CAN signals have more distinct values than \textit{Dynamic} CAN signals.
Since \textit{Switch} CAN signals only have a few values to represent different states, \textit{Switch} CAN signals have lower ratios of distinct values than \textit{Dynamic} and \textit{Verification} CAN signals.
\textit{Unused} signals have the lowest distinct byte value ratio and distinct bit-block value ratio.

By leveraging these identified features, the proposed ByCAN system achieves enhanced performance and accuracy in distinguishing different CAN signal types.
The comprehensive suite of features at both byte and bit levels captures the distinct characteristics and patterns of each signal type more effectively.
}

\subsection{Data Pre-processing}

The initial phase of ByCAN is the data pre-processing of the collected CAN traces that include a variety of CAN IDs.
The first step is to group CAN frames by CAN IDs due to the fact that CAN frames associated with distinct CAN IDs exhibit unique transmission and signal patterns.
Subsequently, the CAN frame's data payload is segmented at byte level and bit level, respectively.
This conversion facilitates the subsequent stages of the byte-level and bit-level CAN signal slicing and labeling.

{\color{black}
\subsubsection{CAN Frame Grouping}
The standard CAN frame has the start of frame, arbitration field, control field, data field, CRC field, ACK field, and end of frame as illustrated in Fig.~\ref{CANframe}.
In practice, CAN frames are recorded and collected by CAN bus analyzers\footnote{CAN bus analyzer can be used to view and log received and transmitted frames from the CAN bus. \url{https://www.influxbigdata.in/post/can-bus-analyzer-choose-the-right-one}} as CAN trace.
Except for the essential data payload of the CAN frame, additional information, such as timestamps, CAN bus channel, and frame state, is covered in the recorded CAN trace by the CAN bus analyzer.
The timestamp is crucial for analyzing the signal patterns over time.
All active functions of the targeted vehicle are represented in the data payloads of CAN frames, where CAN signals vary across occupied bytes and value ranges. 
Different CAN IDs have different transmission patterns and CAN signal patterns for their CAN frames.  
For example, the data payload of CAN ID $\mathrm{0x094}$ could occupy $8$ bytes for active signals, whereas the data payload of CAN ID $\mathrm{0x01AB}$ may only have $3$ bytes used by CAN signals.
The first step of the data pre-processing procedure is to group CAN frames in the recorded CAN trace by their respective CAN ID $C$ before slicing CAN signals within CAN frames.
}

{\color{black}
\subsubsection{Payload Reformatting}
The collected CAN trace is split into multiple CAN traces for different CAN IDs. 
The trace of CAN ID $C$ has $T_C$ rows of CAN frames.
{\color{black}The trace of CAN ID $C$ is further converted into two traces, i.e., $M_C$ and $m_C$, from byte-level and bit-level perspectives, respectively. 
In $M_C$, a 64-bit CAN frame's data payload is converted into eight numerical values, with each byte represented by one value.
The bit-level trace $m_C$ retains the 64-bit representation of the CAN frame's data payload.
Different from existing works, the byte-level trace $M_C$ is introduced to support our new byte-level processing, e.g., byte-level clustering.
The bit-level trace $m_C$ is used in the subsequent bit-level processing, e.g., bit-level slicing.
}
}

\begin{table}
  \centering
  \caption{Notations}
  \begin{tabularx}{\linewidth}{ l X l }
    \toprule
    \thead{\textbf{Notation}} & \thead{\textbf{Description}}\\
    \midrule
    \makecell[lt]{$M_C$/$m_C$ } & CAN trace whose frame payload is segmented by byte/bit for CAN ID $C$ \\
    \addlinespace[5pt]
    \makecell[lt]{$T_C$ } & The number of entries in $M_C$ or $m_C$\\
    \addlinespace[5pt]
    \makecell[lt]{$\Gamma_i^C$} & $i$-th byte of the data payload in the CAN frame for CAN ID $C$ \\
    \addlinespace[5pt]
    \makecell[lt]{$\Gamma_{i,i+1}^C$} & Sliced byte block with the $i$-th and $(i+1)$-th byte in the data payload of the CAN frame for CAN ID $C$ \\
    \addlinespace[5pt]
    \makecell[lt]{$\gamma_k^C$} & $k$-th bit of the data payload in the CAN frame for CAN ID $C$ \\
    \addlinespace[5pt]
    \makecell[lt]{$\gamma_{mn}^C$} & Sliced CAN signal from $m$-th to $n$-th bit of the data payload in the CAN frame for CAN ID $C$  \\
    \addlinespace[5pt]
    \makecell[lt]{$A_i^C$/$a_k^C$/$a_{mn}^C$ } & Average byte/bit/bit-block value of $\Gamma_i^C$/$\gamma_k^C$/$\gamma_{mn}^C$\\
    \addlinespace[5pt]
    \makecell[lt]{$B_i^C$/$b_k^C$/$b_{mn}^C$} & Flip rate of $\Gamma_i^C$/$\gamma_k^C$/$\gamma_{mn}^C$ \\
    \addlinespace[5pt]
    \makecell[lt]{$V_i^C$/$v_{mn}^C$}  & Distinct value set of $\Gamma_i^C$/$\gamma_{mn}^C$\\
    \addlinespace[5pt]
    \makecell[lt]{$U_i^C$/$u_{mn}^C$} & Distinct value ratio of $\Gamma_i^C$/$\gamma_{mn}^C$  \\
    \addlinespace[5pt]
    \makecell[lt]{$\theta_{mn}^C$ } & Labeling parameter of $\gamma_{mn}^C$\\
    \addlinespace[5pt]
    \makecell[lt]{$g$ } & Bit-level labeling function\\
    \addlinespace[5pt]
    \makecell[lt]{$\beta$/$\phi$ } & Flip rate function of $\Gamma_i^C$/$\gamma_{mn}^C$   \\
    \addlinespace[5pt]
    \makecell[lt]{$\zeta$/ $\varpi$ } & Slicing accuracy/coverage\\
    \addlinespace[5pt]
    \makecell[lt]{$\xi$} & Labeling accuracy\\  
    \bottomrule
  \end{tabularx}
  \label{denotion}
\end{table}

\subsection{Signal Slicing}
According to the identified CAN signal Pattern 1, the signal slicing is required to extract different CAN signals from the CAN frames.
CAN signal boundaries are identified through a two-step process: the byte-level signal clustering and the bit-level signal slicing.
Initially, the byte-level features extracted from CAN frames are used to cluster CAN signals at the byte level, establishing preliminary signal delimiters. 
Then, within each byte-level cluster, the boundaries of CAN signals are further sliced and defined at the bit level.
The notations used are detailed in Table~\ref{denotion}. 
Algorithm~\ref{alg:slice} illustrates the CAN signal slicing at both byte and bit level.

{\color{black}
\subsubsection{Byte-level Signal Clustering}
The processed CAN trace $M_C$ of CAN ID $C$ whose data payload segmented by bytes is used in this step.
The byte-level features are extracted from the trace $M_C$ for each byte in the data payload.
$\Gamma_i^C$, $1 \leq i \leq 8$, represents the $i$-th byte of the data payload of the CAN frame for CAN ID $C$.
}

According to the identified Patterns 2 and 3, CAN signals can be firstly located at the byte level to enhance the slicing performance.
ByCAN proposes the new byte-level features, i.e., the byte flip rate $B_i^C$, the average byte value $A_i^C$, and the distinct byte value ratio $U_i^C$.
$B_i^C$ represents the frequency at which the value of $\Gamma_i^C$ changes.
$A_i^C$ is the mean value of $\Gamma_i^C$ and $U_i^C$ indicates the quantity of different values for $\Gamma_i^C$.
Different CAN signals distinguish from each other with different features of $B_i^C$, $A_i^C$, and $U_i^C$.

The byte flip rate of the $\Gamma_i^C$, denoted by $B_i^C$, as given by
\begin{equation}
\label{equ:flipbyte}
\begin{split}
B_i^C&=\frac{\sum_{j=1}^{T_{C}-1} \beta(M_C, i, j)}{T_{C}-1},\\
\beta(M_C, i, j) &= \begin{cases}
    0 ,      & \quad \text{if } \Gamma_{i(j+1)} = \Gamma_{ij},\\
    1 , & \quad \text{if } \Gamma_{i(j+1)} \not= \Gamma_{ij} ,
  \end{cases}
\end{split}
\end{equation}
where $\Gamma_{ij}$ is the value of $\Gamma_i^C$ in the $j$-th entry of $M_C$, $1 \leq j \leq T_C$. 
The byte flip function $\beta(M_C, i, j)$ is $0$ when $\Gamma_i^C$ in adjacent entries are the same and $1$ otherwise.  
We can have $0 \leq B_i^C \leq 1$.

The average value of $\Gamma_i^C$, denoted by $A_i^C$, can be given by
\begin{equation}
\label{equ:byteavg}
A_i^C=\frac{\sum_{j=1}^{T_C} \Gamma_{ij}}{T_C} .
\end{equation}

The distinct byte value ratio of $\Gamma_i^C$, denoted by $U_i^C$, can be calculated by
\begin{equation}
\label{equ:byteunique}
U_i^C=\frac{\left|V_i^C\right|}{256}, \quad 1 \le \left|V_i^C\right| \le 256,
\end{equation}
where $V_i^C$ is the set of distinct values of $\Gamma_i^C$. $\left|V_i^C\right|$ gives the number of elements in $V_i^C$. The maximum number of $\left|V_i^C\right|$ for a byte is $2^8 = 256$, since one byte has 8 bits.

The proposed byte-level features are utilized to cluster CAN signals at byte level by applying clustering algorithms as shown in Algorithm~\ref{alg:slice} line 2.
The number of CAN signals varies for different CAN ID and is unknown, which means the system cannot determine the number of clusters when slicing CAN signals in a CAN frame either at byte or bit level.
Thus, ByCAN utilizes the DBSCAN clustering algorithm that does not require to specify the number of clusters in advance.
Based on the observation of DBC files, CAN signals take up to two bytes position of a CAN frame.
ByCAN restricts the cluster to be maximum two byte long and identifies byte blocks $\Gamma_{i,i+1}^C$ with two bytes and $\Gamma_{i}^C$ with one byte.

\begin{algorithm}[t]
\caption{Slicing algorithm}\label{alg:slice}
\SetAlgoLined
\SetKwInOut{Input}{Input}
\SetKwInOut{Output}{Output}
\Input{CAN trace $M_C$ and $m_C$ for CAN ID $C$}
\Output{Sliced CAN signal $\gamma_{mn}^C$ }
\tcc{\scriptsize Identify byte-level features of $M_C$.}
Compute byte-level features $B_i^C$, $A_i^C$, and $U_i^C$; \qquad  $\triangleright$ with~\eqref{equ:flipbyte},~\eqref{equ:byteavg} and~\eqref{equ:byteunique} \\

\tcc{\scriptsize Cluster byte blocks of $M_C$ by applying the clustering algorithms.}

$\Gamma_{i}^{C} \text{ or } \Gamma_{i,i+1}^{C} \gets \textbf{DBSCAN\_Cluster}(M_C, B_i^C, A_i^C, U_i^C)$\;

\tcc{\scriptsize Slice bit-level signal boundary within byte-level blocks.}

 \For{${\gamma_k^C}$ in $\Gamma_i^C$ or $\Gamma_{i,i+1}^C$}{
 
 Compute the bit-flip rate $b_k^C$; \qquad  $\triangleright$ with~\eqref{equ:flipbit}\;
 Compute the average bit value $a_k^C$; \qquad  $\triangleright$ with~\eqref{equ:bitavg}\;
 
 }
 \tcc{\scriptsize Cluster bit blocks of  $\Gamma_i^C$ or $\Gamma_{i,i+1}^C$ with $m_C$ by applying the clustering algorithms.}

\Return $\gamma_{mn}^{C} \gets \textbf{DBSCAN\_Cluster}(m_C, b_k^C, a_k^C)$;

\end{algorithm}

\subsubsection{Bit-level Signal Slicing}
\label{testtest}
The proposed system slices the bit-level CAN signal boundaries within the identified byte-level signal clusters with $8$ or $16$ bits.

The $k$-th bit of the payload in the CAN frame of CAN ID $C$, denoted by ${\gamma_k^C}$, has $1 \leq k \leq 64$. 
The bit-flip rate of $\gamma_k^C$ is denoted by $b_k^C$, as given by
\begin{equation}
\label{equ:flipbit}
b_k^C=\frac{\sum_{j=1}^{T_C-1} \gamma_{k(j+1)}^i \oplus \gamma_{kj}^i}{T_C-1},\\
\end{equation}
where ${k}$ indicates
the $k$-th bit of the payload in the CAN frame of CAN ID $C$. The ${j}$ denotes the $j$-th CAN frame of ${m_C}$, $1 \leq j \leq T_C$. $\gamma_{kj}^i$ is the value of $\gamma_k^C$ in the $j$-th entry of $m_C$.

The average bit value of ${\gamma_k^C}$, denoted by ${a_k^C}$, is as given by
\begin{equation}
\label{equ:bitavg}
a_k^C=\frac{\sum_{j=1}^{T_C} \gamma_{kj}^i}{T_C} .
\end{equation}

The bit-flip rate $b_k^C$ and average bit value $a_k^C$ are used to cluster bit-level blocks ${\gamma_{mn}^C}$, $m \le n $.
The $m$ and $n$ of ${\gamma_{mn}^C}$ are the start and the end bit positions in the data payload of the CAN frame for CAN ID $C$, respectively. 
Note that $m=n$ when $\gamma_{mn}$ takes only one bit.

{\color{black}
After identifying the byte-level and bit-level features, CAN signal clusters at both levels are generated using the DBSCAN clustering algorithm, as outlined in Algorithm~\ref{alg:slice}.
Under the clustered byte blocks, bit-level features are further calculated to cluster CAN signals at the bit level.
The DBSCAN algorithm is particularly suited for the clustering process because it adeptly handles the variations in signal density, which is common in CAN frames. 
Additionally, the DBSCAN algorithm is suited for clustering signals by each CAN ID or each byte, as it does not require prior knowledge of the number of clusters---a common scenario in reverse engineering the data payloads of CAN frames.
}

\subsection{Signal Labeling}
The signal labeling is illustrated in Algorithm~\ref{alg:label}, including the general labeling and template matching process.
Bit-level features of sliced CAN signal blocks from the Signal Slicing step are identified to associate the sliced CAN signals with the general labels.
{\color{black}
Due to the unknown numbers of CAN signal types, the DBSCAN algorithm is used to cluster based on labeling thresholds, distinguishing different CAN signals' labels using the identified bit-level features.
For the identified \textit{Dynamic} CAN signals, the DTW template matching algorithm is further used to align them with the templates of the OBD-II diagnostic data. 
DTW effectively compensates for discrepancies, such as time differences, shifts, or offsets between the CAN signals and the corresponding templates of the OBD-II diagnostic data.
This alignment is crucial, as CAN frames typically transmit faster than OBD-II diagnostic responses that lack a fixed correspondence ratio with CAN frames. 
}

\subsubsection{General Labeling}
The proposed system labels CAN signals as \textit{Unused}, \textit{Switch}, \textit{Dynamic} and  \textit{Verification}.
{\color{black}
The flip rate is the common parameter to distinguish and label different CAN signal types.
The proposed system improves the labeling process by introducing the labeling parameter and combining it with the flip rate to label CAN signals.
To distinguish the signals with similar flip rates but different ratios of distinct values (e.g., \textit{Unused} and \textit{Switch}), ByCAN combines the flip rate and the ratio of distinct values into a new labeling parameter.
}

The bit-block flip rate of ${\gamma_{mn}^C}$, denoted as $b_{mn}^C$, presents the flip rate of sliced CAN signal blocks in the payload of a CAN frame for CAN ID $C$, as given by
\begin{equation}
\label{equ:bitmnFlip}
\begin{split}
b_{mn}^C&=\frac{\sum_{j=1}^{T_C-1} \phi(m_C, i, {mn}, j)}{T_{C}-1};\\
\phi(m_C, i, {mn}, j) &= \begin{cases}
    0 ,      & \quad \text{if } \gamma_{mn^(j+1)}^i = \gamma_{mn^j}^i;\\
    1 , & \quad \text{if } \gamma_{mn^(j+1)}^i \not= \gamma_{mn^j}^i,
  \end{cases}
\end{split}
\end{equation}
where the value of $\gamma_{mn}^C$ of the $j$-th entry of $m_C$ is denoted by $\gamma_{mn^j}^i$, $1 \leq j \leq T_C$. The bit flip function $\phi$ is $1$ when $\gamma_{mn}^C$ in adjacent entries are different and $0$ otherwise.

The average bit-block value of ${\gamma_{mn}^C}$, denoted by ${a_{mn}}$, can be calculated by
\begin{equation}
\label{equ:bitavg_block}
a_{mn}^C=\frac{\sum_{j=1}^{T_C} \gamma_{{mn}^j}^i}{T_C}.
\end{equation}

The distinct bit-block value ratio of ${\gamma_{mn}^C}$, denoted by ${u_{mn}^C}$, indicates the frequency of different values of a CAN signal in the CAN trace and is as given by
\begin{equation}
\label{equ:bitunique}
u_{mn}^C=\frac{\left|v_{mn}^C\right|}{2^{n-m+1}}, \quad 1 \le \left|v_{mn}^C\right| \le 2^{n-m+1},
\end{equation}
where ${v_{mn}^C}$ is the set of distinct bit-block values of ${\gamma_{mn}^C}$, and $\left|{v_{mn}^C}\right|$ is the size of the set ${v_{mn}^C}$.
$2^{n-m+1}$ is the maximum number of the bit block taken up from $m$-th bit to $n$-th bit (m $\leq$ n), i.e., the maximum number of $\left|v_{mn}^C\right|$.

\begin{algorithm}[t]
\caption{Labeling algorithm}\label{alg:label}
\SetAlgoLined
\SetKwInOut{Input}{Input}
\SetKwInOut{Output}{Output}
 \Input{Sliced CAN signal ${\gamma_{mn}^C}$, CAN trace $m_C$ for CAN ID $C$, time series of the templates of OBD-II diagnostic data $S_x$}
\Output{CAN signal label of ${\gamma_{mn}^C}$}
\tcc{\scriptsize Identify bit-level features of all $\gamma_{mn}^C$.}
 Compute bit-level features $b_{mn}^C$, $a_{mn}^C$, and $u_{mn}^C$; \qquad  $\triangleright$ with~\eqref{equ:bitmnFlip},~\eqref{equ:bitavg_block} and~\eqref{equ:bitunique} \\

Compute labeling parameter $\theta_{mn}^C$; \qquad  $\triangleright$ with~\eqref{equ:bitmag}\\

\tcc{\scriptsize Determine labeling thresholds ${\varepsilon_0}$ by applying clustering algorithms with $\theta_{mn}^C$, $b_{mn}^C$ and $a_{mn}^C$ of all bit-blocks.}

${\varepsilon_0}$ $ \gets$ \textbf{Cluster}($\theta_{mn}^C$, $b_{mn}^C$, $a_{mn}^C$)\;

\ForEach {$\gamma_{mn}^C$}{

\tcc{\scriptsize Assign general labels.}
Label $\gamma_{mn}^C$ with $g(\theta_{mn}^C, b_{mn}^C)$; \qquad  $\triangleright$ with~\eqref{equ:bitfunc}\\
\tcc{\scriptsize Assign descriptive labels to Dynamic signals by using the template matching algorithms.}
\If{$g(\theta_{mn}^C, b_{mn}^C)$ = $Dynamic$}{
        \tcc{\scriptsize Convert the CAN message values of $\gamma_{mn}^C$ into the time seires sequence $E_x$.}
        $E_x \gets$ \textbf{Serialize}(${\gamma_{mn}^C}$)\;
        ${\gamma_{mn}^C}$.\textbf{TemplateMatch}($S_x$, $E_x$)\;
    }

}

\Return Labels of $\gamma_{mn}^C$\;
\end{algorithm}

The labeling parameter ${\theta_{mn}^C}$ is defined to label the sliced CAN signals ${\gamma_{mn}^C}$, as given by
\begin{equation}
\label{equ:bitmag}
\theta_{mn}^C =  b_{mn}^C \times u_{mn}^C,
\end{equation}
{\color{black}where $0 \leq \theta_{mn}^C \leq 1$.
The parameter ${\theta_{mn}^C}$ combines $b_{mn}^C$ and $u_{mn}^C$ to consider both the bit-block flip rate and the ratio of distinct bit-block values when labeling CAN signals, thereby maximizing the dissimilarity of different labels.
Instead of using $u_{mn}^C$  only to label CAN signals, ${\theta_{mn}^C}$ can distinguish CAN signals with similar flip rates and different ratios of distinct values.}

{\color{black}
\textit{Unused} CAN signal blocks have ${\theta_{mn}^C=0}$ and $b_{mn}^C=0$ by which the signal can be deduced as unchanged no matter the value it represents.
\textit{Verification}, \textit{Switch} and \textit{Dynamic} CAN signals have $b_{mn}^C>0$. 
\textit{Switch} CAN signals are triggered less frequently than \textit{Dynamic} CAN signals. 
\textit{Dynamic} and \textit{Verification} CAN signals usually have higher $b_{mn}^C$.
Given all calculated ${\theta_{mn}^C}$, the threshold ${\varepsilon_0}$ is determined by unsupervised clustering algorithms and used to label \textit{Switch}, \textit{Dynamic}, and \textit{Verification} CAN signals.  
The bit-level labeling function ${g}$ is given by
}

{\color{black}
\begin{equation}
\small
\label{equ:bitfunc}
g(\theta_{mn}^C, b_{mn}^C) = \begin{cases}
    Unused  & \! \text{if } \theta_{mn}^C = 0 \quad \text{and} \quad b_{mn}^C = 0 ;\\
    Switch  & \! \text{if } 0 < \theta_{mn}^C  \le \varepsilon_{0};\\
    Dynamic   & \! \text{if } \varepsilon_{0} \leq \theta_{mn}^C  \quad \text{and} \quad b_{mn}^C < 0.99 ;\\
    Verification  & \! \text{if } \varepsilon_{0} \leq \theta_{mn}^C  \quad \text{and} \quad b_{mn}^C \geq 0.99  .
  \end{cases}
\end{equation}
The labeling function $g$ takes into account ${\theta_{mn}^C}$ and $b_{mn}^C$ to match the label based on different CAN signals' features. \textit{Unused} CAN signals have both ${\theta_{mn}^C}$ and $b_{mn}^C$ that are equal to 0. The \textit{Switch} CAN signals have lower ${\theta_{mn}^C}$ than \textit{Dynamic} and \textit{Verification} CAN signals. The \textit{Verification} CAN signals have higher ${\theta_{mn}^C}$ and the highest $b_{mn}^C$.
The proposed system sets the threshold value for $b_{mn}^C$ at 0.99 to distinguish between \textit{Verification} and \textit{Dynamic} CAN signals.
This threshold mitigates the uncommon scenario where \textit{Verification} CAN signals fail to toggle in each CAN frame within the recorded CAN trace.
}

{\color{black}
\subsubsection{Template Matching}
OBD-II diagnostic data are used in this step since the information of some vehicle states, e.g., vehicle speed and engine speed, can be extracted from OBD-II diagnostic data.
ByCAN sends OBD-II diagnostic requests, such as the request of PID 0C for the engine speed state, to target vehicle.
Following the public OBD-II diagnostic data standard\footnote{The standard OBD-II PIDs are defined by SAE J1979, which contains the formula to translate the OBD-II diagnostic response into meaningful data. \url{https://en.wikipedia.org/wiki/OBD-II_PIDs}}, the requested OBD-II diagnostic data are examined to translate and extract the values of the relevant states.
The sliced and labeled \textit{Dynamic} CAN signals are then matched with the extracted information from the OBD diagnostic response to further determine the descriptive signal labels without prior knowledge by leveraging template matching algorithms. 
The proposed system builds the matching template with the collected OBD-II diagnostic data that contain the real-time dynamic signal values, such as Vehicle Speed and Accelerator Pedal Position.
ByCAN associates sliced CAN signals with descriptive labels: Speed, Wheel Angle, Throttle Pedal Position, and Brake Pedal Position Related.
}

The generated templates represent the time series of corresponding signal values. 
$S_x$ denotes a time series template of the OBD-II diagnostic data for one \textit{Dynamic} CAN signal $x$, where $S_x = \{s_1, ...,s_a,...,s_y\}$. 
ByCAN measures the similarity between the sliced \textit{Dynamic} CAN signal candidates and the templates. 
The time series sequence of a sliced \textit{Dynamic} CAN signal candidate is denoted as $E_x$, $E_x = \{e_1, ...,e_b,...,e_z\}$.
Note that $z = T_C$. 
The values of $E_x$ are the values of the labeled signal block $\gamma_{mn}$ of $m_C$ in sequence. 
The defined time interval of the OBD-II diagnostic data is different from the transition rate of different CAN IDs. 
Thus, length $y$ of $S_x$ may be longer or shorter than the length $z$ of $E_x$. 

In the proposed system, the DTW algorithm is used as the similarity measure since $S_x$ and $E_x$ are unequal-length time series~\cite{berndt1994using}. 
A $w$-by-$z$ matrix is constructed to align $S_x$ and $E_x$ using~\eqref{equ:ab}, where $d_{a, b}$ is the element ($a$, $b$) in the constructed $w$-by-$z$ matrix to present the Euclidean Distance between the points $s_a$ and $e_b$ {\color{black}(i.e., one for the \textit{Dynamic} CAN signal and the other for the template of the OBD-II diagnostic data). }

\begin{equation}
\label{equ:ab}
d_{a, b}=\left(s_{a}-e_{b}\right)^{2}.
\end{equation}

The warping path $W$ is given in~\eqref{equ:W}, where the $k$-th element $w_k = (a,b)_k$. 
\begin{equation}
\label{equ:W}
\begin{split}
W_x = \{w_1, w_2, ...,w_k,...,w_K\}, \\
\max(y,z)\le K < y+z-1 .
\end{split}
\end{equation}

The optimal path between $S_x$ and $E_x$ is determined by finding the minimum path from the set of all admissible paths, as given by 
\begin{equation}
\label{equ:DTW}
DTW(S_x, E_x)= \min \left( \sqrt{\sum_{k=1}^{K} w_{k}} \right),
\end{equation}
{\color{black}where $DTW(S_x, E_x)$ is the optimal path (i.e., the shortest path in our case) for the \textit{Dynamic} CAN signal candidate with all templates of the OBD-II diagnostic data.} The proposed system measures the optimal paths for the signal candidate with all assessed OBD-II diagnostic data templates. The descriptive label with the shortest path is finally associated with the signal candidate.

\subsection{Assessment Criteria}

To quantify the performance of ByCAN, we propose three metrics, i.e., slicing accuracy $\zeta$, slicing coverage $\varpi$, and labeling accuracy $\xi$.

\smallskip
\noindent\textbf{Slicing accuracy ($\zeta) $}
 indicates the percentage of correctly sliced CAN signal bits, is as given by
\begin{equation}
\label{equ:Acc}
\zeta = \sum_{id=1}^{x} \frac{ S_{id}^{\gamma_{mn}^C}}{ G_{id}^{\gamma_{mn}^C}},
\end{equation}
where $S_{id}^{\gamma_{mn}^C}$ is the number of bits that are correctly deduced CAN signal boundaries for $\gamma_{mn}^C$ of the $id$-th CAN ID frame.
$G_{id}^{\gamma_{mn}^C}$ is the number bits of exact signal boundaries for $\gamma_{mn}^C$. 
The $x$ standards for the number of CAN IDs analyzed in the system evaluation.

\smallskip
\noindent\textbf{Slicing coverage ($\varpi$)}
represents the portion of sliced CAN signals that are correctly located within the ground-truth CAN signal boundaries.
$\varpi$ can be calculated by
\begin{equation}
\label{equ:Coverage}
\varpi = \sum_{id=1}^{x} \frac{ \mathbb{S}_{id}^{\gamma_{mn}^C}}{ G_{id}^{\gamma_{mn}^C}},
\end{equation}
where $x$ is the number of CAN IDs, the $\mathbb{S}_{id}^{\gamma_{mn}^C}$ stands for the number of CAN signals that are located within the ground-truth CAN signal boundaries of the $id$-th CAN ID signal.

\smallskip
\noindent\textbf{Labeling accuracy ($\xi$)}
stands for the system performance of labeling CAN signals with the general or the descriptive labels.
$\xi $ assesses the sliced CAN signals that are correctly labeled, as given by 
\begin{equation}
\label{equ:CoverageLabel}
\xi = \sum_{id=1}^{x} \frac{ L_{id}^{\gamma_{mn}^C}}{ \mathbb{G}_{id}^{\gamma_{mn}^C}},
\end{equation}
where $L_{id}^{\gamma_{mn}^C}$ is the number of correct labeled CAN signals for $\gamma_{mn}^C$ of the $id$-th CAN ID frame.
The label of $\gamma_{mn}$ in the ground truth is determined by the label that takes up more bits in $\gamma_{mn}$.
$\mathbb{G}_{id}^{\gamma_{mn}^C}$ is the ground-truth label of $\gamma_{mn}$.

\section{Evaluation and Results}
\label{performance}

{\color{black}
\subsection{Experimental Setup}
In our experiments, the real-world CAN traces were collected from three cars of different makes and models: the 2006 Mazda 3 Automatic, the 2017 Honda Civic Manual, and the 2022 Toyota RAV4 Hybrid. 
The PEAK PCAN-USB Pro\footnote{PEAK PCAN-USB Pro is one of CAN analyzers bridging a PC with CAN network to collect CAN frames. \url{https://www.peak-system.com/PCAN-USB-Pro-FD.366.0.html?&L=1}} adapter was used as the CAN analyzer in the experiment to listen and record CAN frames on CAN bus via the OBD-II port.
An OBD-II cable was connected between the OBD-II port of the target vehicle and the PEAK PCAN-USB Pro, and the PEAK PCAN-USB Pro was connected with the USB port of the experiment laptop to run the PCAN-Explorer 6\footnote{PEAK PCAN-Explorer 6 is the Windows system software compatible with the PEAK PCAN-USB Pro to receive, send, and record CAN frames. \url{https://www.peak-system.com/PCAN-Explorer-6.415.0.html?&L=1}} to record CAN frames into trace files.
The CAN frames were collected passively from the CAN bus, while the OBD-II diagnostic data were collected by sending related OBD-II requests.
OBD-II disgnostic requests with the PIDs of 04 (calculated engine load), 0C (engine speed), 0D (vehicle speed), 11 (throttle position), 45 (relative throttle position), 47 (absolute throttle position B), 48 (absolute throttle position C), 49 (absolute throttle position D), 4A (absolute throttle position E), and 4B (absolute throttle position F) were sent to the target vehicle.
Both CAN frames and OBD-II data were monitored and recorded into the trace files simultaneously.

The collected CAN trace files were converted into CSV files using the compatible software PEAK-Converter\footnote{PEAK PCAN-Converter is the Windows system software compatible with the PEAK PCAN-USB Pro to convert trace files to various output formats for the processing and analysis purpose. \url{https://www.peak-system.com/PEAK-Converter.554.0.html?&L=1}}.
The parsed CSV files of the collected CAN trace files were fed into the ByCAN system.
The byte-level features, i.e., the flip rate, the average value, the distinct byte value ratio of byte, were extracted from the payload of recorded frames.
ByCAN restricts the byte clusters to have a minimum length of 1~byte and a maximum length of 2 bytes.
Within the identified byte clusters, the bit blocks of CAN signals were further clustered with the DBSCAN clustering algorithm with the bit-level features.
We ran the data collection and static analysis with the DELL Latitude 5420 laptop, with 16 GB RAM and an 11th Gen Intel(R) Core(TM) i7-1185G7 CPU.

All possible car functions, e.g., acceleration, deceleration, wheel steering and car indicators, were triggered to collect as many CAN signals as we could.
The data collection step takes 2 minutes on average to collect the regular CAN frames and the OBD-II diagnostic data for ByCAN to reverse engineer CAN signals.
The experimental results were then assessed with the ground truth from the OpenDBC repository~\cite{openDBC}.
Since the DBC files from the OpenDBC were generated from manual reverse engineering, not all CAN signals in the tested cars were covered in the DBC files from OpenDBC.
Thus, only the CAN signals exist on both the collected CAN trace files and the ground-truth DBC files were evaluated. 
The DBC files utilized in our experiments were cross-validated with laboratory vehicles by engineers at the Insurance Australia Group (IAG) Research Center, ensuring the reliability of our ground truth data.
The evaluation focused on the system's slicing and labeling performance, examining the impact of CAN signal type, signal length, CAN IDs and the number of CAN frames.
The CAN traces collected from individual cars were used as the input to the proposed system for different rounds of evaluations.
The overview of the experimental results is the average performance of all three cars tested in our experiments, which is evaluated and discussed in this section.
}

\begin{table}[t]
  \centering
  \caption{{\color{black}Comparison of overall performance}}
  \label{tab:overall}
  \begin{tabular}
  {p{0.25\linewidth}p{0.13\linewidth}p{0.15\linewidth}p{0.15\linewidth}}
    \toprule[1pt]
      \textbf{System} & \textbf{Slicing accuracy $\zeta$} & \textbf{Slicing coverage $\varpi$} & \textbf{Labeling accuracy $\xi$} \\
      \midrule[1pt]
      \multirow{1}{*}{\textbf{ByCAN}}  
      & \textbf{80.21\%} & \textbf{95.21\% } &  \textbf{68.72\%} \\
      \midrule
      \multirow{1}{*}{CAN-D~\cite{Verma}}
      & 63.88\% &  89.21\%  &  67.62\%\\
       \midrule
      \multirow{1}{*}{READ~\cite{Marchetti}}  
      & 51.99\% &  83.61\%  &  64.48\% \\
      \bottomrule[1pt]
    \end{tabular}
\end{table}

{\color{black}
\subsection{Overall Performance}
The proposed system ByCAN is compared against two other RE systems, i.e., READ~\cite{Marchetti} and CAN-D~\cite{Verma}, with the evaluation metrics comprising slicing accuracy ($\zeta$), slicing coverage ($\varpi$), and labeling accuracy ($\xi$), as summarized in Table~\ref{tab:overall}.
The READ system relies solely on the bit-flip rate to slice and label CAN signals. 
The CAN-D system leverages multiple bit-level features, including the bit-flip rate, two-bit distributions, and entropy. 
The proposed ByCAN system distinguishes the READ and CAN-D from the initial step of byte-level clustering before identifying bit-level signal boundaries within the byte clusters.
The experimental results presented are the averaged performance metrics of evaluated systems with real-world CAN traces collected from all tested cars included in our study.

In our experiments, ByCAN demonstrates a superior slicing accuracy $\zeta$ of 80.21\%, which notably surpasses that of CAN-D and READ, with slicing accuracy $\zeta$ of 63.88\% and 51.99\%, respectively.
ByCAN achieves slicing coverage $\varpi$ of 95.21\%, which also outperforms the CAN-D (89.21\%) and READ (83.61\%).
In terms of labeling accuracy $\xi$, ByCAN maintains the lead with a 68.72\% labeling accuracy, while CAN-D and READ are behind with 67.62\% and 64.48\%, respectively.
The experimental result indicates that the proposed ByCAN system gains a better overall performance in decoding CAN signal specifications without any prior knowledge.
}

\begin{figure*}[!htbp]
    \centering
    \subfigure[]{
    \begin{minipage}[t]{0.4\textwidth}
    \centering
    \includegraphics[width=3in]{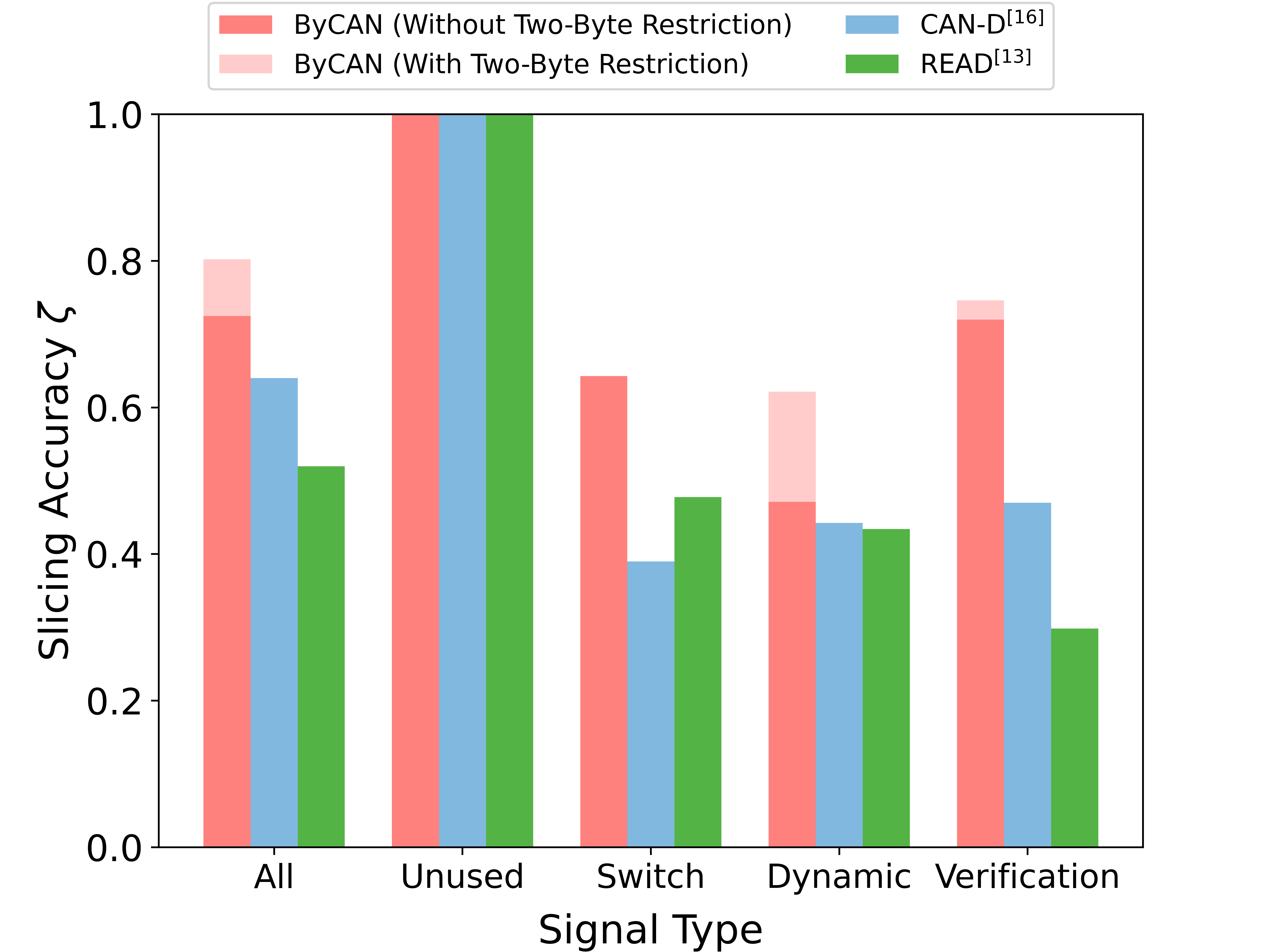}
    \end{minipage}
    \label{fig:signalType_acc}
    }
    \subfigure[]{
    \begin{minipage}[t]{0.4\textwidth}
    \centering
    \includegraphics[width=3in]{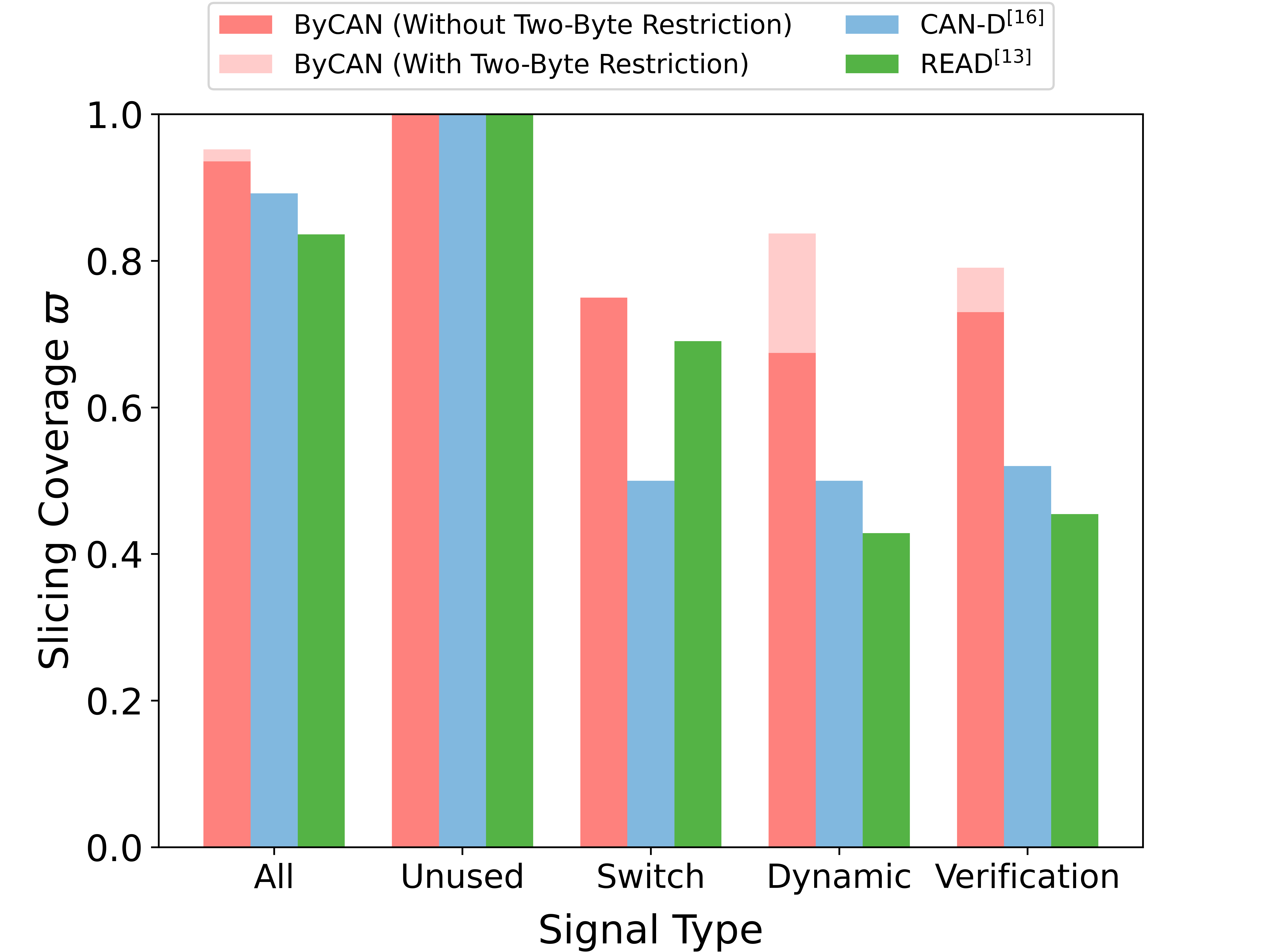}
    \end{minipage}
    \label{fig:signalType_cov}
    }
    \subfigure[]{
    \begin{minipage}[t]{0.4\textwidth}
    \centering
    \includegraphics[width=3in]{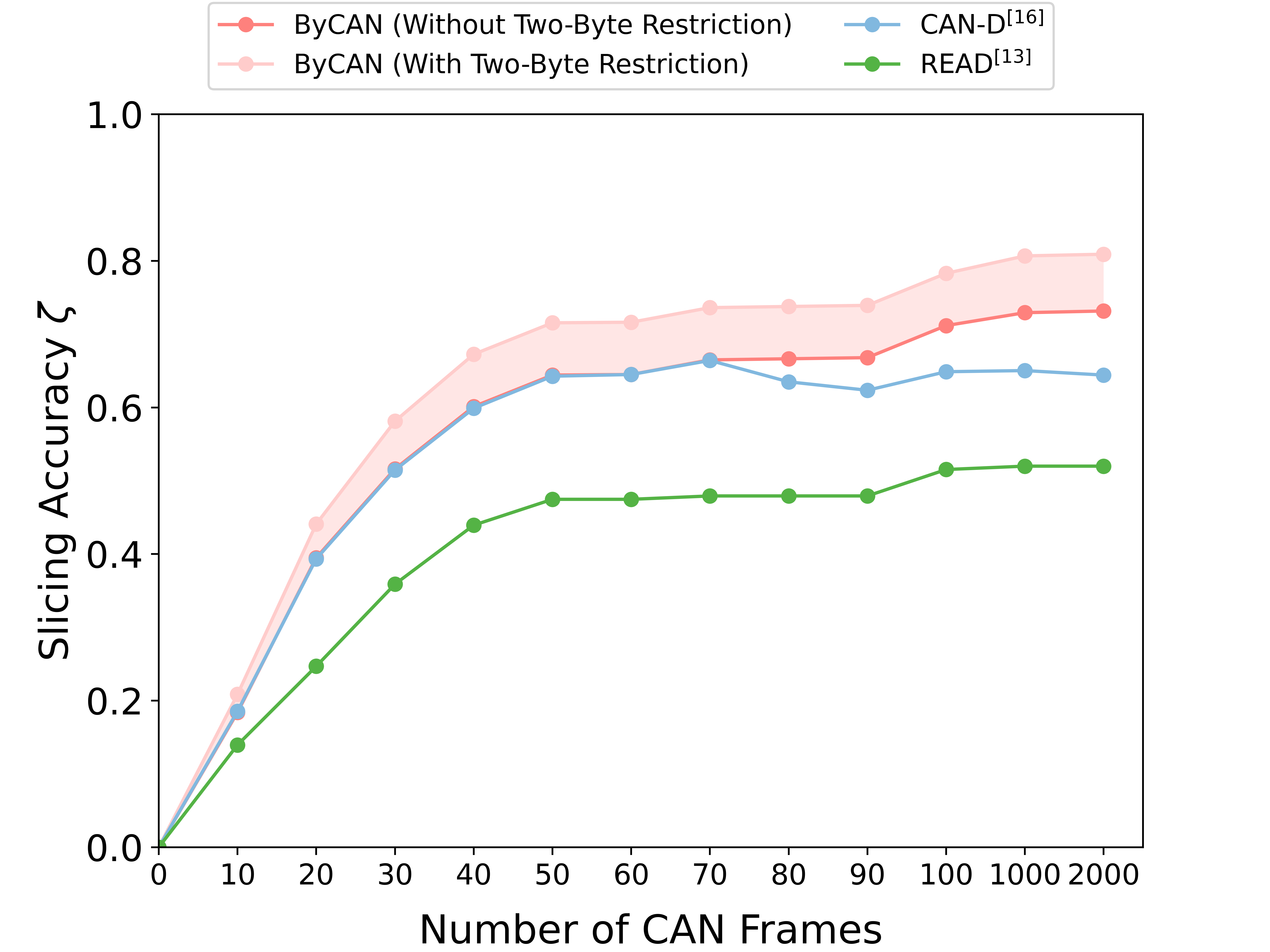}
    \end{minipage}
    \label{fig:msgLen_acc}
    }
     \subfigure[]{
    \begin{minipage}[t]{0.4\textwidth}
    \centering
    \includegraphics[width=3in]{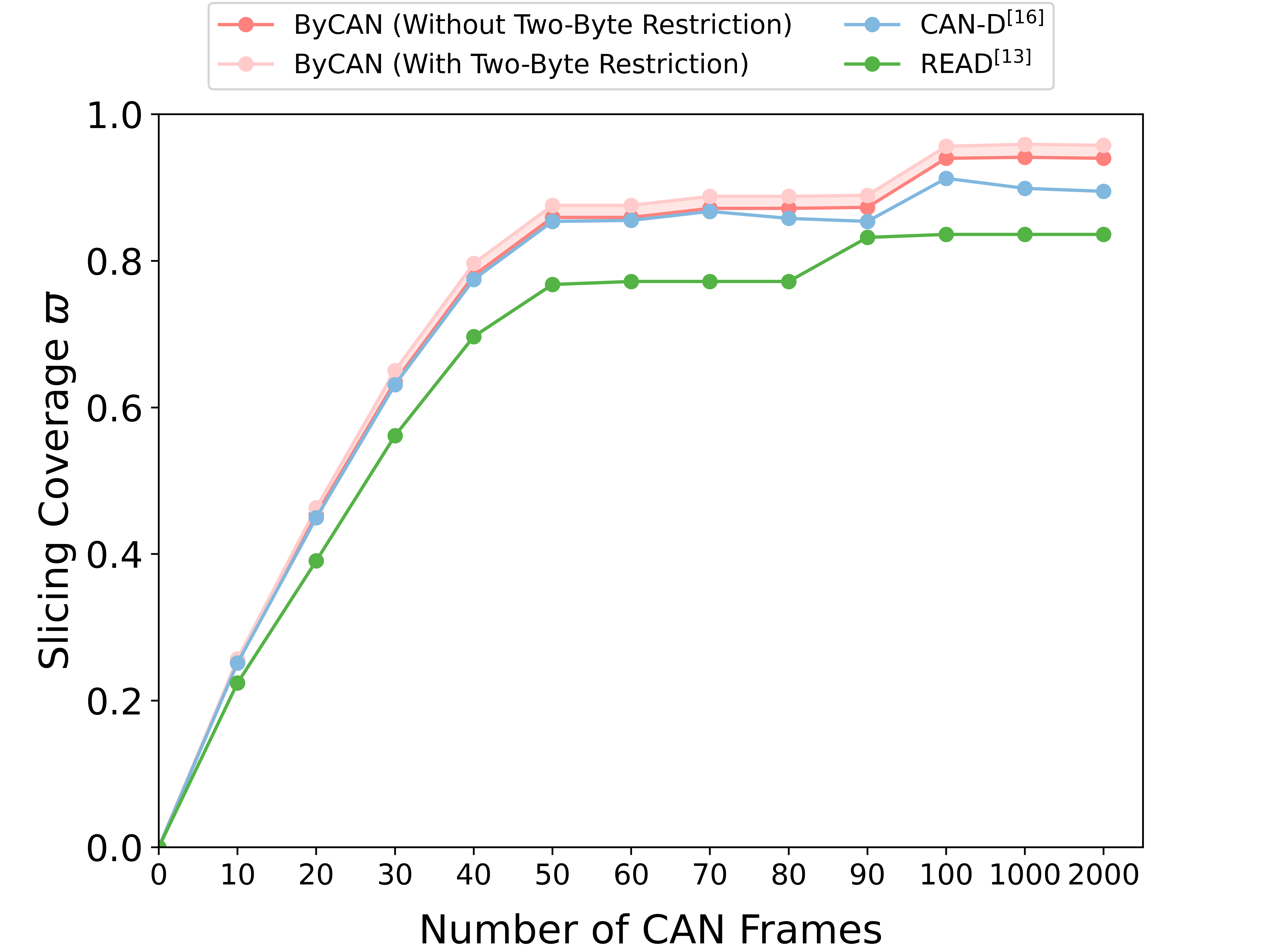}
    \end{minipage}
    \label{fig:msgLen_cov}
    }
\caption{\textcolor{black}{Comparison of slicing accuracy and slicing coverage of different systems: The $y$-axis is the slicing accuracy $\zeta$ and the $x$-axis is CAN signal type in (a); the $y$-axis is the slicing coverage $\varpi$ and the $x$-axis is CAN signal type in (b);  the $y$-axis is the slicing accuracy $\zeta$ and the $x$-axis is the number of CAN frames in (c); the $y$-axis is the slicing coverage $\varpi$ and the $x$-axis is the number of CAN frames in (d). Note that the \textit{Verification} CAN signals represent both the \textit{Counter} and \textit{Checksum} signals. }}
\label{fig:SliceAcc}
\end{figure*}

\subsection{Slicing Performance}
\label{slicePerformance}

{\color{black}
\subsubsection{Size of Byte-level Clusters} 
According to Pattern 2 and our observation, most CAN signals take less than 16 bits in the CAN frames' data payloads.
We compare the ByCAN's performance with and without the two-byte restriction in the byte-level clustering step, and ByCAN with the restriction has better overall performance, as shown in Fig.~\ref{fig:SliceAcc}.
The setting of the two-byte restriction for the byte-level clusters' length enhances the overall slicing performance of ByCAN from $72.47\%$ to $80.21\%$ for slicing accuracy $\zeta$, and from $93.58\%$ to $95.22\%$ for slicing coverage $\varpi$, as demonstrated in Figs.~\ref{fig:signalType_acc} and~\ref{fig:signalType_cov}.
ByCAN with two-byte restriction also gains better slicing accuracy for \textit{Dynamic} and \textit{Verification} CAN signals, which is $15.03\%$ and $2.62\%$ higher than that without the two-byte restriction.
In terms of the slicing coverage $\varpi$, the setting of two-byte restriction for the signal clustering increases $\varpi$ from $67.44\%$ to $83.72\%$ for \textit{Dynamic} CAN signal, and from $73\%$ to $79.07\%$ for \textit{Verification} CAN signal.
The reason is that the restriction of two-byte on the signal cluster's length allocates CAN signals more accurately at the byte-level boundary.
Since the \textit{Verification} and \textit{Dynamic} CAN signals tend to take one or two bytes, the restriction setting improves the relevant slicing accuracy and coverage.
As shown in Figs.~\ref{fig:msgLen_acc} and~\ref{fig:msgLen_cov}, the ByCAN system with two-byte restriction also has better slicing accuracy and coverage than that without two-byte restriction across different numbers of CAN frames.
}

\subsubsection{CAN Signal Type}

When comparing the ByCAN system with two-byte restriction to the CAN-D and READ systems, ByCAN demonstrates higher slicing accuracy and coverage than the other two systems, regardless of the signal type, as depicted in Figs.~\ref{fig:signalType_acc} and~\ref{fig:signalType_cov}.
ByCAN reaches up to a slicing accuracy $\zeta$ of $80.21\%$ that is $16.21\%$ higher than the CAN-D and $28.22\%$ higher than the READ, as shown in Fig.~\ref{fig:signalType_acc}.
The same trend can be found in Fig.~\ref{fig:signalType_cov} where ByCAN has the slicing coverage $\varpi$ of $93.58\%$ and the CAN-D and READ systems have that of $89.21\%$ and $83.61\%$.
ByCAN is superior in slicing the \textit{Switch}, \textit{Dynamic} and \textit{Verification} (i.e., \textit{Counter} and \textit{Checksum}) CAN signals as indicated in Figs.~\ref{fig:signalType_acc} and~\ref{fig:signalType_cov}.
The reason is that the \textit{Switch} signals with similar functionalities and \textit{Verification} signals tend to be located within one or two bytes, where the byte-level clustering of ByCAN enhances the slicing performance by avoiding the over-slicing at the bit level.
Taking the advantage of byte-level clustering, ByCAN has a better slicing coverage $\varpi$ than the CAN-D and READ systems for \textit{Dynamic} CAN signals that usually take one or two bytes long.
For \textit{Unused} CAN signals, all systems can reach the $100\%$ slicing accuracy and coverage because \textit{Unused} signals have a bit-flip rate that always equals $0$.

\begin{table*}[!htbp]
  \centering
  \caption{\textcolor{black}{Comparison of slicing accuracy and coverage statistics across CAN IDs}}
  \label{tab:performance_CANID}
  \begin{threeparttable}
  \begin{tabular}
  {p{0.12\linewidth}p{0.1\linewidth}p{0.07\linewidth}p{0.07\linewidth}p{0.07\linewidth}p{0.07\linewidth}p{0.07\linewidth}p{0.15\linewidth}}
    \toprule[1pt]
      \textbf{} & \textbf{System} & \textbf{Min} & \textbf{Max} & \textbf{Mean}  & \textbf{Median}& \textbf{Variance} & \textbf{Standard Deviation} \\
      \midrule[1pt]
      \multirow{4}{*}{Slicing accuracy $\zeta$}  
      & \textbf{ByCAN} &  \textbf{48.43\% } &  \textbf{96.43\%} &  \textbf{80.93\%} &  \textbf{83.59\%}  & \textbf{0.0211}  &   \textbf{0.1453}  \\
      & CAN-D~\cite{Verma} &  9.38\%  &  96.43\% &  66.65\%  &  80.73\% &  0.0949 &   0.3081 \\
      & READ~\cite{Marchetti} &   6.25\%  & 95.31\%   &   54.47\%  &   64.84\% &   0.1098  &  0.3314\\
      \midrule
      \multirow{4}{*}{Slicing coverage $\varpi$} 
        & \textbf{ByCAN} & \textbf{72.73\%}  &  \textbf{100\%} & \textbf{94.48\%}  &  \textbf{96.41\%}  & \textbf{0.0053}  &  \textbf{0.0730}  \\
      & CAN-D~\cite{Verma} &  27.27\%  &  100\% &  81.19\%  &  91.86\% &  0.0538 &   0.2320 \\
      & READ~\cite{Marchetti} &   20.00\%  & 98.31\%   &   72.47\%  &   84.53\% &   0.0646   &  0.2541\\
      \bottomrule[1pt]
    \end{tabular}
   \end{threeparttable}
\end{table*}

\begin{table*}[!htbp]
  \centering
  \caption{Comparison of slicing accuracy and coverage of CAN Signal length}
  \label{tab:performance_signalLength}
  \begin{threeparttable}
  \begin{tabular}
  {p{0.12\linewidth}p{0.12\linewidth}p{0.08\linewidth}p{0.08\linewidth}p{0.08\linewidth}p{0.08\linewidth}p{0.08\linewidth}p{0.08\linewidth}}
    \toprule[1pt]
      \textbf{} & \textbf{System} & \textbf{1 Bit$^1$} & \textbf{2 Bits} & \textbf{4 Bits}  & \textbf{8 Bits} & \textbf{10 Bits} & \textbf{16 Bits}\\
      \midrule[1pt]
      \multirow{3}{*}{Slicing accuracy $\zeta$}  
      & ByCAN & \textbf{77.50\% } & 41.94\% & \textbf{81.52\%} & \textbf{45.63\% }& \textbf{70\%}&  \textbf{59.38\%} \\
      & CAN-D~\cite{Verma} & 57.50\%  & 37.10\% & 56.52\%  & 27.5\% & 35\% &  56.25\% \\
      & READ~\cite{Marchetti} & 50\%  & \textbf{83.87\% }& 30.43\%  & 25\% & 50\%  & 43.75\% \\
      \midrule
      \multirow{3}{*}{Slicing coverage $\varpi$} 
      & ByCAN & \textbf{70\%}  & 48.39\% & \textbf{100\%} & \textbf{65\%} & \textbf{100\%} & \textbf{81.25\%} \\
      & CAN-D~\cite{Verma} & 52.50\%  & 41.94\% & 73.91\% & 35\% &  50\% &  56.25\% \\
      & READ~\cite{Marchetti} & 50\%  & \textbf{83.87\%} & 30.43\% & 25\% & 50\% &  43.75\% \\
      \bottomrule[1pt]
    \end{tabular}
         \begin{tablenotes}
  \footnotesize
  \item[1] Exclude \textit{Unused} CAN signals since they comprise the largest portion of the 1-bit long signal. Each \textit{Unused} signal is regarded as occupying only one bit position, even when they appear consecutively within a CAN frame.
  \end{tablenotes}
   \end{threeparttable}
\end{table*}

{\color{black}
\subsubsection{Number of CAN Frames}
As shown in Figs.~\ref{fig:msgLen_acc} and~\ref{fig:msgLen_cov}, all systems accurately slice a comparable number of CAN signals.
Both $\zeta$ and $\varpi$ increase with the number of CAN frames fed into the RE systems.
However, the slicing performance converges when the number of CAN frames exceeds $1,000$, a point at which most functionalities are triggered in our experiments.
The optimum value of $T_C$ is considered as $1,000$ for both the slicing accuracy and coverage.
When $T_C = 1,000$, ByCAN with two-byte restriction ($\zeta$ = $80.90\%$) surpasses the CAN-D system ($\zeta$ = $65.03\%$) and the READ system ($\zeta$ = $51.99\%$), as demonstrated in Fig.~\ref{fig:msgLen_acc}.
Although the slicing accuracy differs significantly among the systems, there is only a slight variation in the slicing coverage $\varpi$ among them, as illustrated in Fig.~\ref{fig:msgLen_cov}.
When $T_C = 1,000$, ByCAN with two-byte restriction reaches $95.90\%$ of $\varpi$ while the CAN-D and READ systems achieve $89.89\%$ and $83.61\%$ of $\varpi$, respectively.
}

{\color{black}
\subsubsection{CAN IDs}
As indicated in Table~\ref{tab:performance_CANID}, the performance of systems when slicing CAN signals shows variability among different CAN IDs. 
This variability can be attributed to the differences in the number and types of CAN signals contained within the frames of each CAN ID.
Specifically, ByCAN exhibits slicing accuracy $\zeta$ ranging from 48.43\% to 96.43\%, signifying substantial performance variation tied to the specific CAN ID. 
By contrast, CAN-D and READ display a wider range (9.38\% to 96.43\% and 6.25\% to 95.31\%, respectively) of the slicing accuracy, indicating a broader disparity in performance across CAN IDs.
Furthermore, while ByCAN achieves high slicing coverage $\varpi$ with minimal variability as evidenced by the low variance of 0.0053, it still reveals a significant gap between its minimum and maximum coverage (72.73\% to 100\%). 
The CAN-D and READ systems have a higher variance (0.0538 and 0.0646) than ByCAN, underscoring the inconsistency in the slicing coverage across different CAN IDs.
A notable spread in the slicing coverage is also demonstrated for CAN-D  (27.27\% to 100\%) and READ  (20.00\% to 98.31\%) in Table~\ref{tab:performance_CANID}.
}

\subsubsection{CAN Signal Length}

The system performance of slicing CAN signals with different signal lengths is detailed in Table~\ref{tab:performance_signalLength}.
Given the diverse lengths of CAN signals, this paper specifically focuses the commonly used bit lengths - namely, $1$ bit, $2$ bits, $4$ bits, $8$ bits, $10$ bits and $16$ bits - for the generality.
It is important to note that \textit{Unused} CAN signals, all considered as $1$-bit long, are excluded from this discussion.
ByCAN demonstrates higher $\zeta$ and $\varpi$ for CAN signals with lengths of $1$ bit, $4$ bits, $8$ bits, $10$ bits and $16$ bits.
Notably, ByCAN has better performance for longer CAN signal lengths.
For instance, ByCAN achieves slicing coverage $\varpi$ of $81.25\%$ for $16$-bit long CAN signals, nearly twice the value obtained by the READ system at $43.75\%$ for the same length. 
Specifically, ByCAN achieves a $\zeta$ of $81.52\%$ and $70\%$ for $4$-bit long CAN signals and attains a perfect $\zeta$ and $\varpi$ (both $100\%$) for $10$-bit long CAN signals.

\begin{figure*}[!htbp]
    \centering
    \subfigure[]{
    \begin{minipage}[t]{0.3\textwidth}
    \centering
    \includegraphics[width=2.5in]{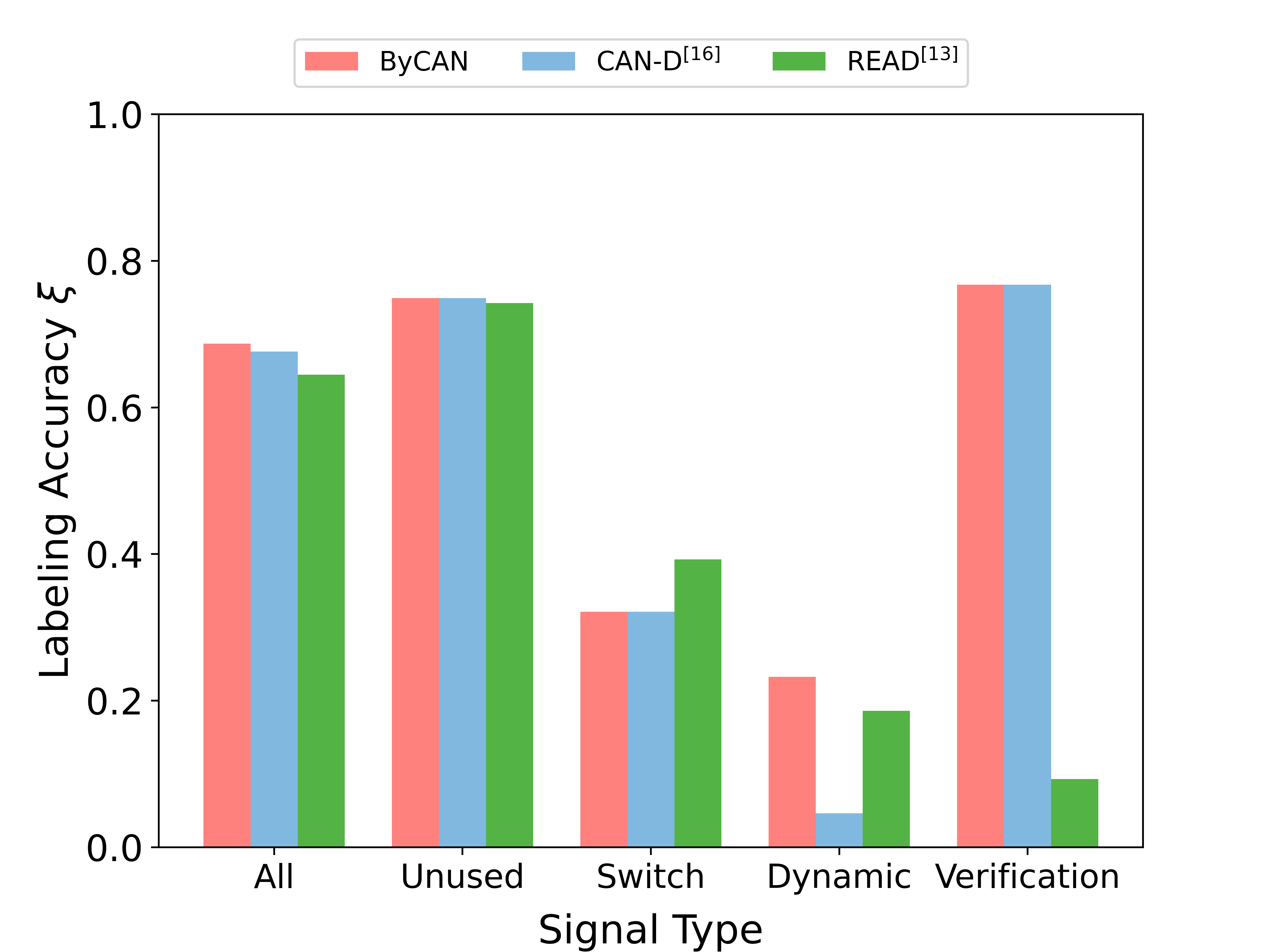}
    \end{minipage}
    \label{fig:signalType_Acc_La}
    }
    \subfigure[]{
    \begin{minipage}[t]{0.3\textwidth}
    \centering
    \includegraphics[width=2.5in]{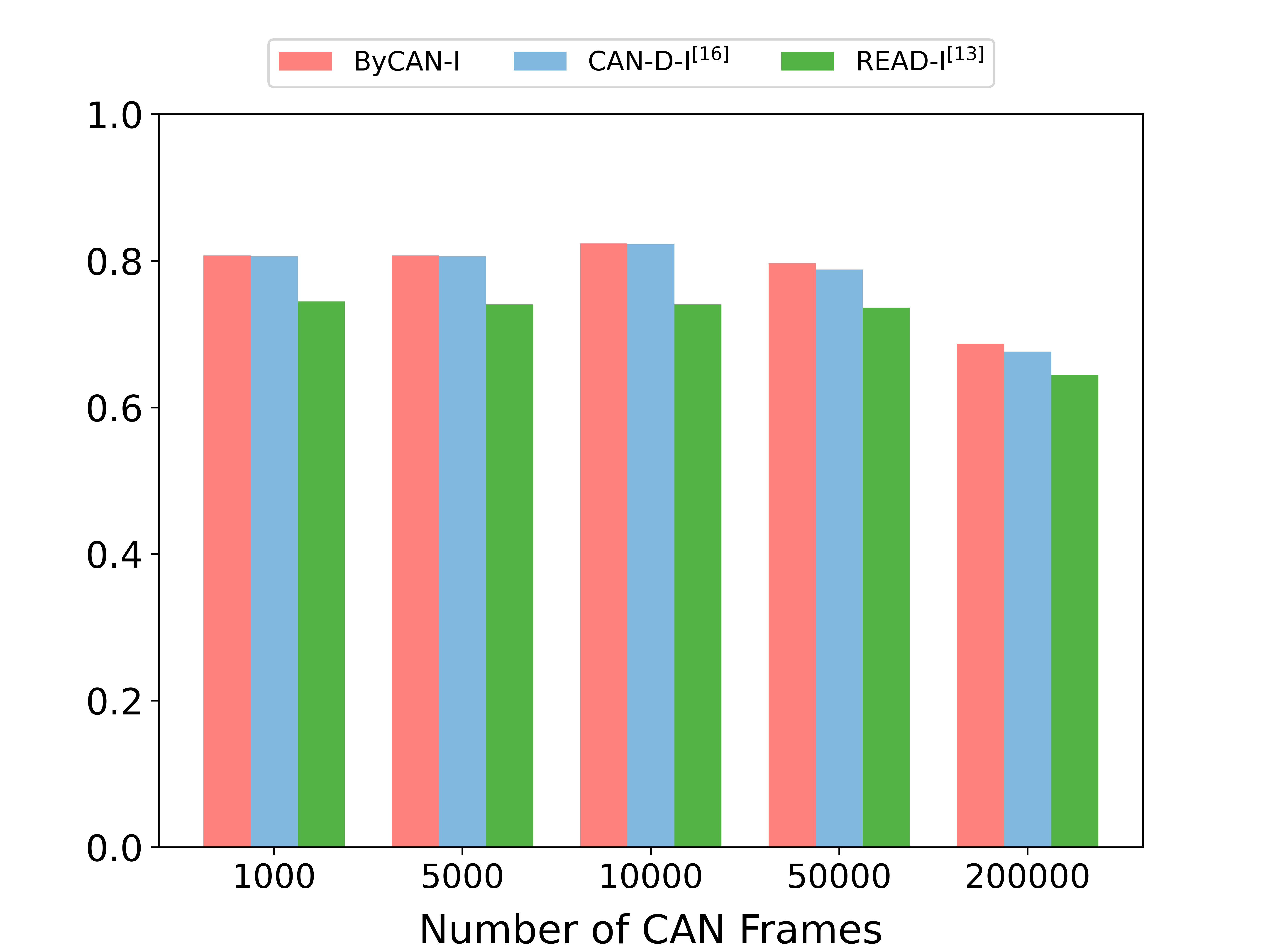}
    \end{minipage}
    \label{fig:msgLen_Acc_La}
    }
     \subfigure[]{
    \begin{minipage}[t]{0.3\textwidth}
    \centering
    \includegraphics[width=2.5in]{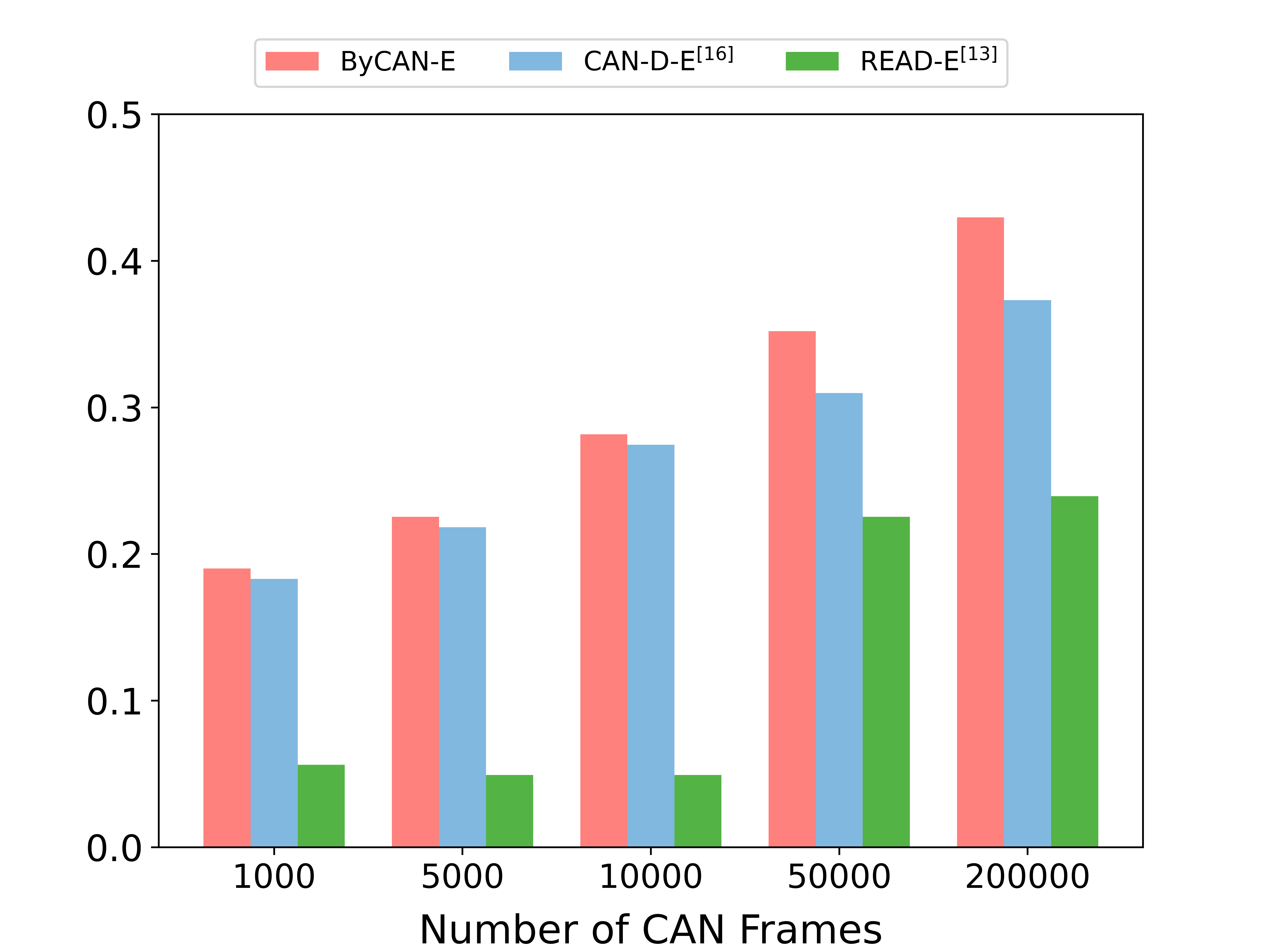}
    \end{minipage}
    \label{fig:msgLen_Acc_La_E}
    }
\caption{\textcolor{black}{Comparison of labeling accuracy of different systems: The $y$-axis is the labeling accuracy $\xi$ for all subplots. The $x$-axis is the CAN signal type in (a), and the $x$-axis is the number of CAN frames in (b) and (c). 
Subplot (b) gives the labeling accuracy including \textit{Unused} CAN signals while subplot (c) gives the labeling accuracy excluding \textit{Unused} CAN signals.
}
}
\label{fig:LabelAcc}
\end{figure*}

\subsection{Labeling Performance}
\label{labelPerformance}

\subsubsection{General Labeling}

The labeling accuracy $\xi$ for the general labels of CAN signals is discussed among different RE systems in Fig.~\ref{fig:LabelAcc}.
As indicated in Fig.~\ref{fig:LabelAcc}, ByCAN system that labels CAN signals with multiple inputs (i.e., labeling threshold and flip rate of signal block) outperforms the READ and CAN-D system which use only bit-flip rate and multiple bit-level features, respectively.
The general labels of CAN signals are \textit{Unused}, \textit{Switch}, \textit{Dynamic}, and \textit{Verification}.
In Fig~\ref{fig:LabelAcc}, the impact of different CAN signal types and the number of CAN frames on $\xi$ are discussed.

The overall labeling accuracy $\xi$ of ByCAN is $1.09\%$ slightly greater than that of CAN-D and $4.23\%$ higher than that of READ.
The performances of the three different RE systems have little difference for labeling the \textit{Unused} CAN signals.
The ByCAN outperforms CAN-D and READ except in labeling the \textit{Switch} CAN signals.
The READ system achieves $39.29\%$ labeling accuracy for the \textit{Switch} CAN signals, while the ByCAN and CAN-D have the same labeling accuracy of $32.14\%$.
READ has higher labeling accuracy for the \textit{Switch} CAN signals due to the excessively sliced small boundaries at the bit level that lead to more CAN signals identified as \textit{Switch}.
For READ, the trade-off of the high labeling accuracy of the \textit{Switch} ($39.29\%$) and \textit{Dynamic} ($18.60\%$) CAN signals is the lower labeling accuracy of the \textit{Verification} ($9.3\%$) CAN signals.
This discrepancy may be due to the READ solely using the bit-flip rate, which could cause mislabeling of \textit{Verification} signals as \textit{Dynamic} or \textit{Switch}.
This mislabeling is particularly likely when the bit in the most significant position (MSB) exhibits a significantly different bit-flip rate compared to the bit in the least significant position (LSB) of the \textit{Verification} signals.

{\color{black}
The impact of the number of CAN frames is demonstrated in Figs.~\ref{fig:msgLen_Acc_La} and~\ref{fig:msgLen_Acc_La_E}.
The labeling accuracy $\xi$ of the assessments, including \textit{Unused} CAN signals, slightly decreases with a larger size of CAN frames as indicated in Fig.~\ref{fig:msgLen_Acc_La}.
When $T_C = 10,000$, the labeling accuracy of the ByCAN system, including \textit{Unused} CAN signals (i.e., ByCAN-I) is $82.38\%$, that of CAN-D (i.e., CAN-D-I) is $82.24\%$, and that of READ (i.e., READ-I) is $74.04\%$.
The labeling accuracy of the ByCAN-I, CAN-D-I and READ-I drops to $68.72\%$, $67.62\%$, and $64.48\%$, respectively.
On the contrary, the labeling accuracy $\xi$ of ByCAN-E, CAN-D-E, and READ-E increases largely from $28.17\%$, $27.46\%$, and $4.93\%$ to $42.96\%$, $37.32\%$, and $22.54\%$ when the number of CAN frames increases from \textcolor{black}{$10,000$} to \textcolor{black}{$200,000$}, as demonstrated in Fig.~\ref{fig:msgLen_Acc_La_E}.
The features of CAN signals become more accurate with more CAN frames, which improves the labeling accuracy for \textit{Switch}, \textit{Dynamic}, and \textit{Verification} CAN signals and reduces the false positive rate of \textit{Unused} CAN signals.
Based on our experiments, the optimum of $T_C$ for the labeling accuracy is $10,000$.
}

Compared to the labeling accuracy in Fig.~\ref{fig:signalType_Acc_La} which counts all CAN signals including those rarely triggered signals during data collection, the labeling accuracy in Table~\ref{tab:performance_switchDym} evaluates the precision of labeling performance for triggered CAN signals. 
The \textit{Switch} and \textit{Dynamic} CAN signal types are highlighted due to the significant differences in labeling accuracy when using triggered signals versus all CAN signals from the ground truth.
Not all \textit{Switch} and \textit{Dynamic} CAN signals can be triggered during the experiments, leading to the low labeling accuracy.
The labeling accuracy increases from $32.14\%$ in in Fig.~\ref{fig:signalType_Acc_La} to $94.74\%$ in Table~\ref{tab:performance_switchDym} for \textit{Switch} signals.
The same trend is found for \textit{Dynamic} CAN signals whose labeling accuracy increases from $23.26\%$ to $70.59\%$.
The result shows a good labeling performance of ByCAN for all triggered and collected CAN signals.

\begin{table}[t]
  \centering
  \caption{Labeling accuracy of triggered signals}
  \label{tab:performance_switchDym}
  \begin{threeparttable}
  \begin{tabular}
  {p{0.4\linewidth}p{0.4\linewidth}}
    \toprule[1pt]
      \textbf{Type of Triggered Signals} & \textbf{Labeling Accuracy $\xi^1$}  \\
      \midrule[1pt]
      \multirow{1}{*}{\textit{Switch}}  
      &  94.74\%   \\
      \midrule
      \multirow{1}{*}{\textit{Dynamic}} 
      &  70.59\% \\
      \bottomrule[1pt]
    \end{tabular}
         \begin{tablenotes}
  \footnotesize
  \item[1] $\xi$ is estimated specifically for the CAN signals that are triggered during the experiments. 
  \end{tablenotes}
   \end{threeparttable}
\end{table}

\subsubsection{Descriptive Labeling}
ByCAN leverages the template matching to find the similarity between sliced CAN signals and the templates of OBD diagnostic data to associate them with descriptive meanings.
In the proposed system, only the \textit{Dynamic} CAN signals are further processed in this step to efficiently identify the descriptive labels with minimum time cost.
ByCAN firstly introduces the DTW algorithm in labeling the descriptive labels to CAN signals with the templates of OBD diagnostic data.
By using DTW, the small time shift problem between the collected CAN frames and the OBD diagnostic data can be solved. 

Within the descriptive labels that can be extracted from the OBD diagnostic data, the Engine Speed, Vehicle Speed, Throttle Position, Fuel Rail Gauge Pressure, and Calculated Engine Load stand out as target labels that have potential significance for subsequent assessments conducted by industry engineers.
Few \textit{Dynamic} CAN signals are sliced due to the limited number of known functions to be triggered.
However, ByCAN still shows a high labeling accuracy of identifying descriptive labels for the sliced \textit{Dynamic} CAN signals.
In our experiments, all signals matched as the Engine Speed and Throttle Position labels are correctly categorized, and the labeling accuracy is drop to $50\%$ for those with the Vehicle Speed label. 
Although ByCAN does not have the same level of labeling accuracy for non-speed-related labels, it is worth noting that the speed-related labels have valuable insights and guidance for future research.

\section{Conclusion}
\label{conclusion}

In this paper, we proposed the fully automated RE system named ByCAN to decode the specification of the Controller Area Network frames without prior knowledge.
Based on the observation of DBC files and CAN frames, CAN signal patterns are identified and used in ByCAN for reverse engineering CAN signals.
ByCAN is the first to introduce multiple byte-level and bit-level features that capture the characteristics and patterns of CAN signals to slice and label CAN signals.
ByCAN applies the byte-level clustering with the DBSCAN algorithm for enhancing the slicing and labeling performance.
The proposed ByCAN system is also the first to use the DTW algorithm in matching sliced \textit{Dynamic} CAN signals with the templates of OBD diagnostic data for associating CAN signals with descriptive labels, facilitating further research.
The experiments validated that ByCAN enhances the performance of CAN signal slicing and labeling with real-world data from various cars.
ByCAN requires only a short CAN trace in our experiments to achieve a good reverse-engineering result.

\section*{Acknowledgment}

We thank our industry partner IAG for providing access to test vehicles. We thank Thanh Phuoc Nguyen for helping with technical issues and experiments. This work is funded in part by the University of Technology Sydney, Insurance Australia Group (IAG), and the iMOVE CRC under Grant 5-028. This work is also supported by the Cooperative Research Centres program, an Australian Government initiative.

\bibliographystyle{IEEEtran}
\bibliography{conference_re}

\vskip -2\baselineskip plus -1fil 

\begin{IEEEbiography}[{\includegraphics[width=1in,height=1.25in,clip,keepaspectratio]{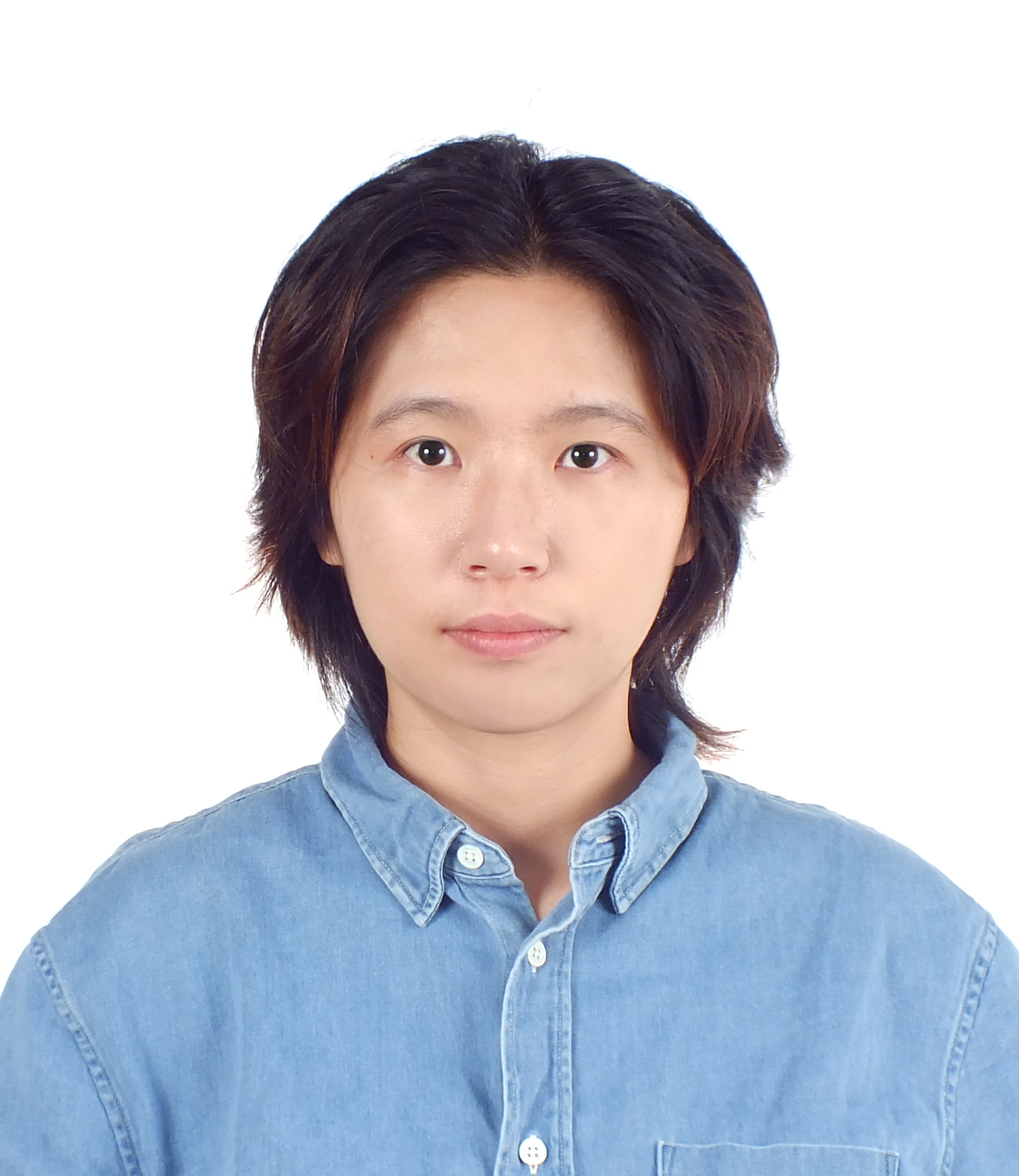}}]{Xiaojie Lin} is currently a Ph.D. student at University of Technology Sydney, Australia. She received the M.I.T. degree in 2020 from the Faculty of Engineering and Information Technology, University of Technology Sydney, Australia. Her main research interests include cybersecurity, privacy, and machine learning.
\end{IEEEbiography}

\vskip -2\baselineskip plus -1fil

\begin{IEEEbiography}[{\includegraphics[width=1in,height=1.25in,clip,keepaspectratio]{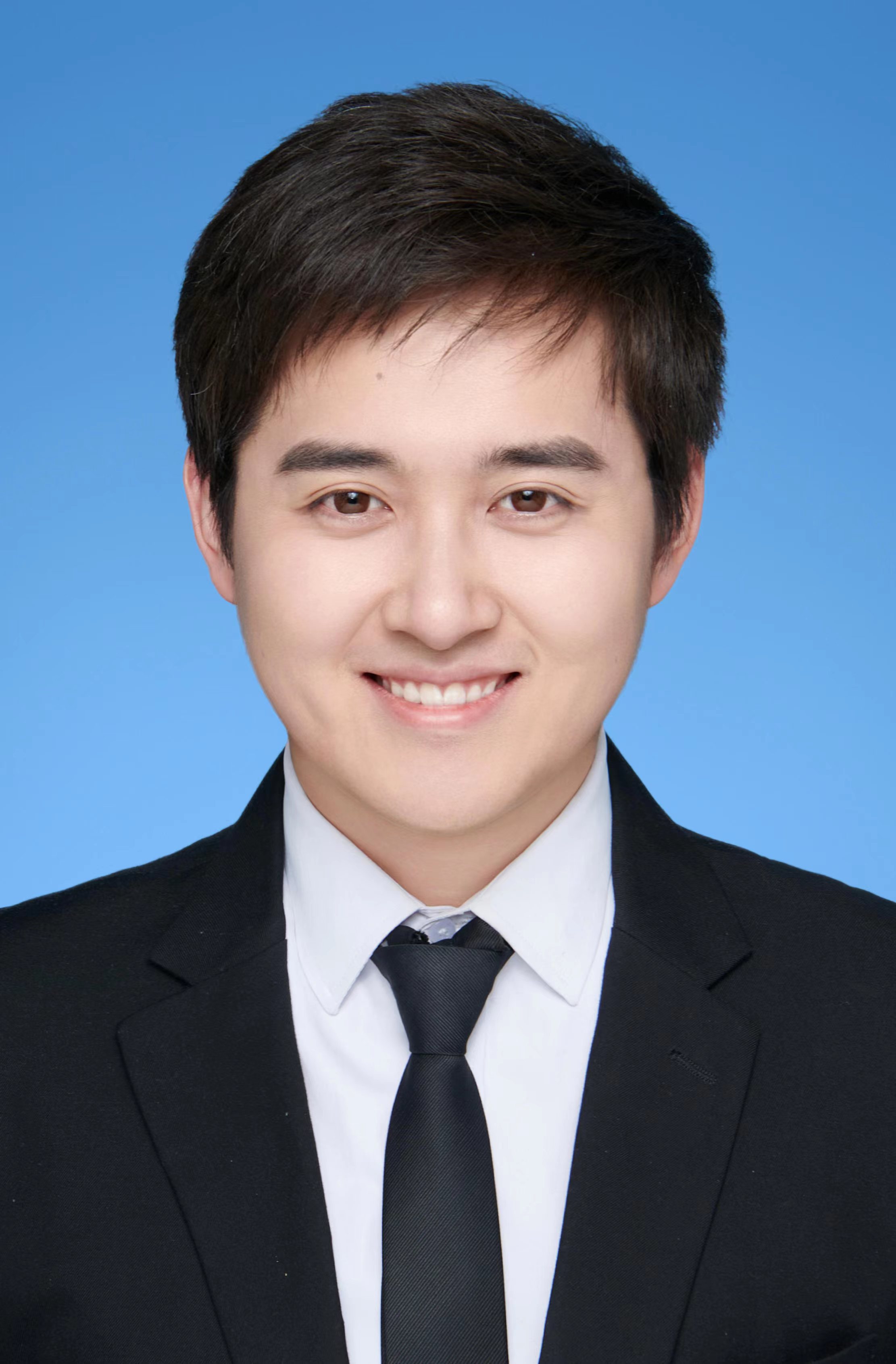}}]{Baihe Ma} received his B.E. and M.E. degrees from Xidian University China in 2016 and 2019, and Ph.D. degrees from University Technology  Sydney, Australia in 2024. His research interests include cybersecurity, data privacy, location privacy, and machine learning.
\end{IEEEbiography}

\vskip -2\baselineskip plus -1fil

\begin{IEEEbiography}[{\includegraphics[width=1in,height=1.25in,clip,keepaspectratio]{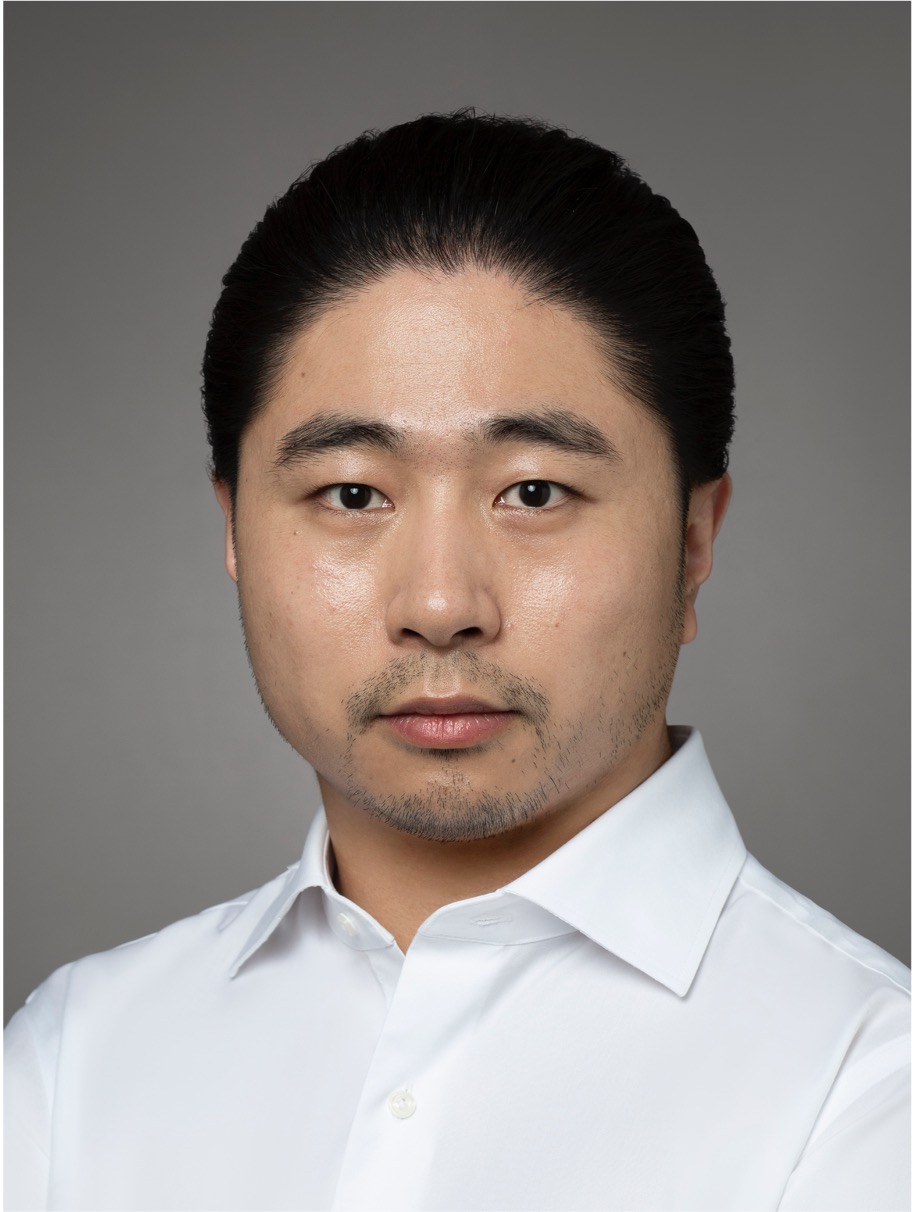}}]{Xu Wang} received his B.E. from Beijing Information Science and Technology University, China, in 2010, and dual Ph.D. degrees from Beijing University of Posts and Telecommunications, China, in 2019 and University of Technology Sydney, Australia, in 2020. He is currently a Senior Lecturer with the School of Electrical and Data Engineering, University of Technology Sydney, Australia. His research interests include cybersecurity, blockchain, privacy, and network dynamics.
\end{IEEEbiography}

\vskip -2\baselineskip plus -1fil

\begin{IEEEbiography}[{\includegraphics[width=1in,height=1.25in,clip,keepaspectratio]{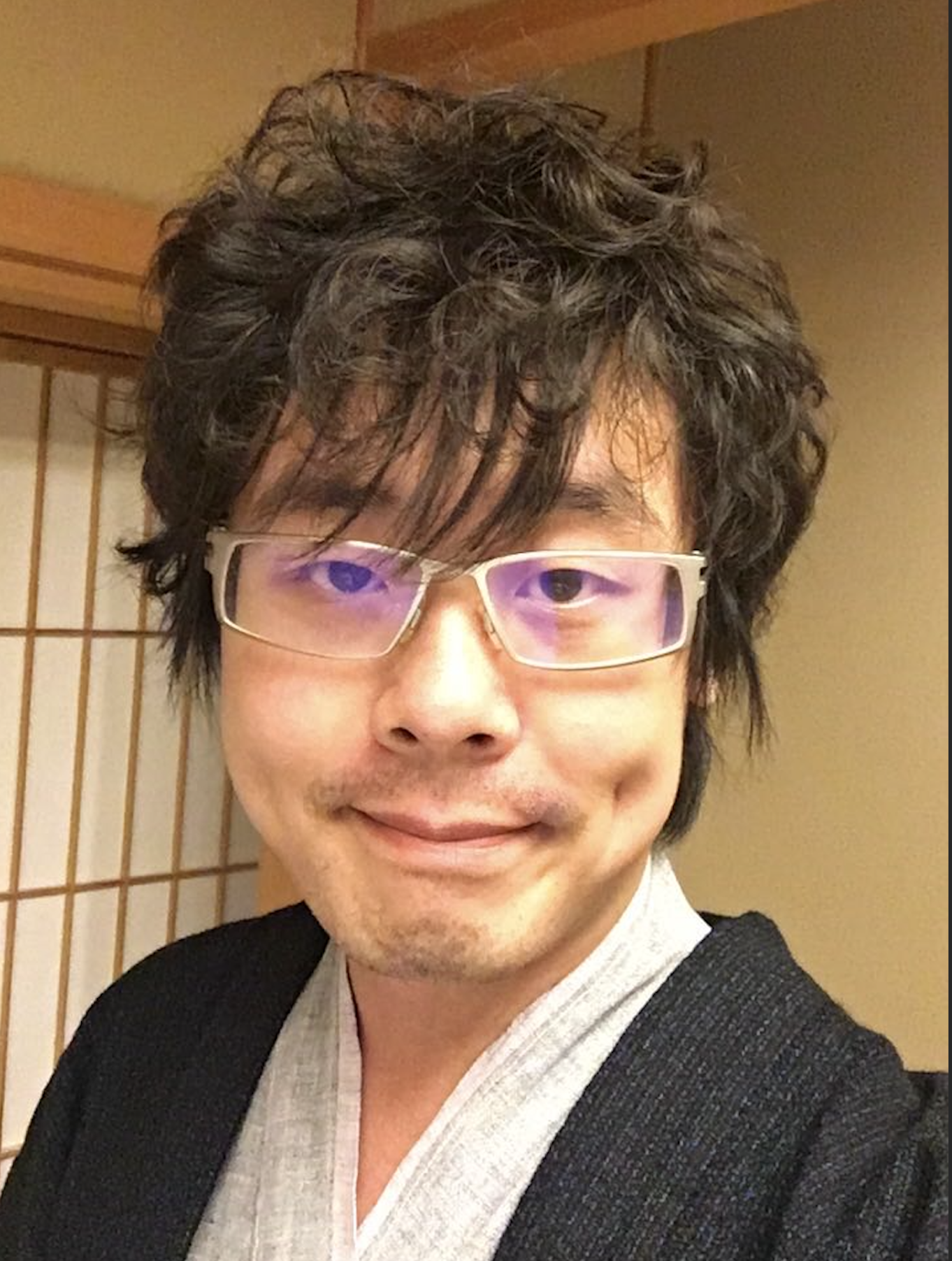}}]{Guangsheng Yu} is currently a Post-Doctoral Research Fellow with CSIRO Data61 and an adjunct lecturer at University of Technology Sydney (UTS). He received the Ph.D. degree in 2021 with the Faculty of Engineering and Information Technology, UTS. He received the B.Sc. degree and M.Sc degree from the University of New South Wales, Sydney, Australia, from 2011 to 2015. His main research interests lie in cybersecurity, blockchain, and distributed learning.
\end{IEEEbiography}

\vskip -2\baselineskip plus -1fil

\begin{IEEEbiography}[{\includegraphics[width=1in,height=1.25in,clip,keepaspectratio]{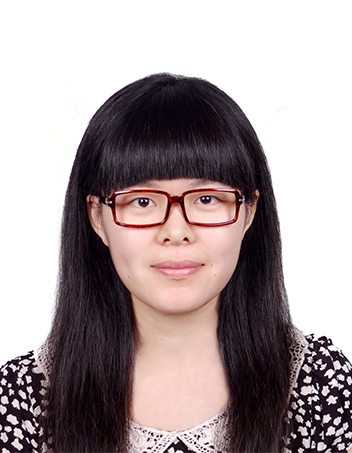}}]{Ying He} received the B.Eng. degree in telecommunications engineering from Beijing University of Posts and Telecommunications, Beijing, China, in 2009, and the Ph.D. degree in telecommunications engineering from the University of Technology Sydney, Australia, in 2017. She is currently a Senior Lecturer with the School of Electrical and Data Engineering, University of Technology Sydney. Her research interests are physical layer algorithms in wireless communication with machine learning, vehicular communication, spectrum sharing and satellite communication.
\end{IEEEbiography}

\vskip -2\baselineskip plus -1fil

\begin{IEEEbiography}[{\includegraphics[width=1in,height=1.25in,clip,keepaspectratio]{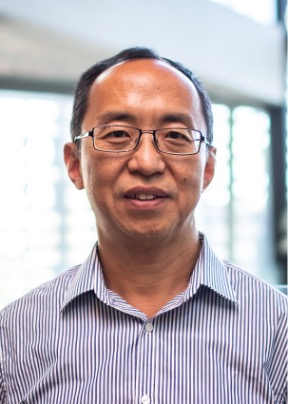}}]{Ren Ping Liu} (M’09-SM’14) received his B.E. degree from Beijing University of Posts and Telecommunications, China, and the Ph.D. degree from the University of Newcastle, Australia, in 1985 and 1996, respectively. He is a Professor and Head of Discipline of Network \& Cybersecurity at University of Technology Sydney (UTS). As a research leader, a certified network professional, and a full stack web developer, he has delivered networking and cybersecurity solutions to government agencies and industry customers. His research interests include wireless networking, 5G, IoT, Vehicular Networks, 6G, Cybersecurity, and Blockchain. He has supervised over 30 PhD students and has over 200 research publications. Professor Liu was the winner of NSW iAwards 2020 for leading the BeFAQT (Blockchain-enabled Fish provenance And Quality Tracking) project. He was awarded the Australian Engineering Innovation Award 2012 and the CSIRO Chairman’s Medal for contributing to the Wireless Backhaul project. Professor Liu was the founding chair of the IEEE NSW VTS Chapter and a Senior Member of IEEE.
\end{IEEEbiography}

\vskip -2\baselineskip plus -1fil

\begin{IEEEbiography}[{\includegraphics[width=1in,height=1.25in,clip,keepaspectratio]{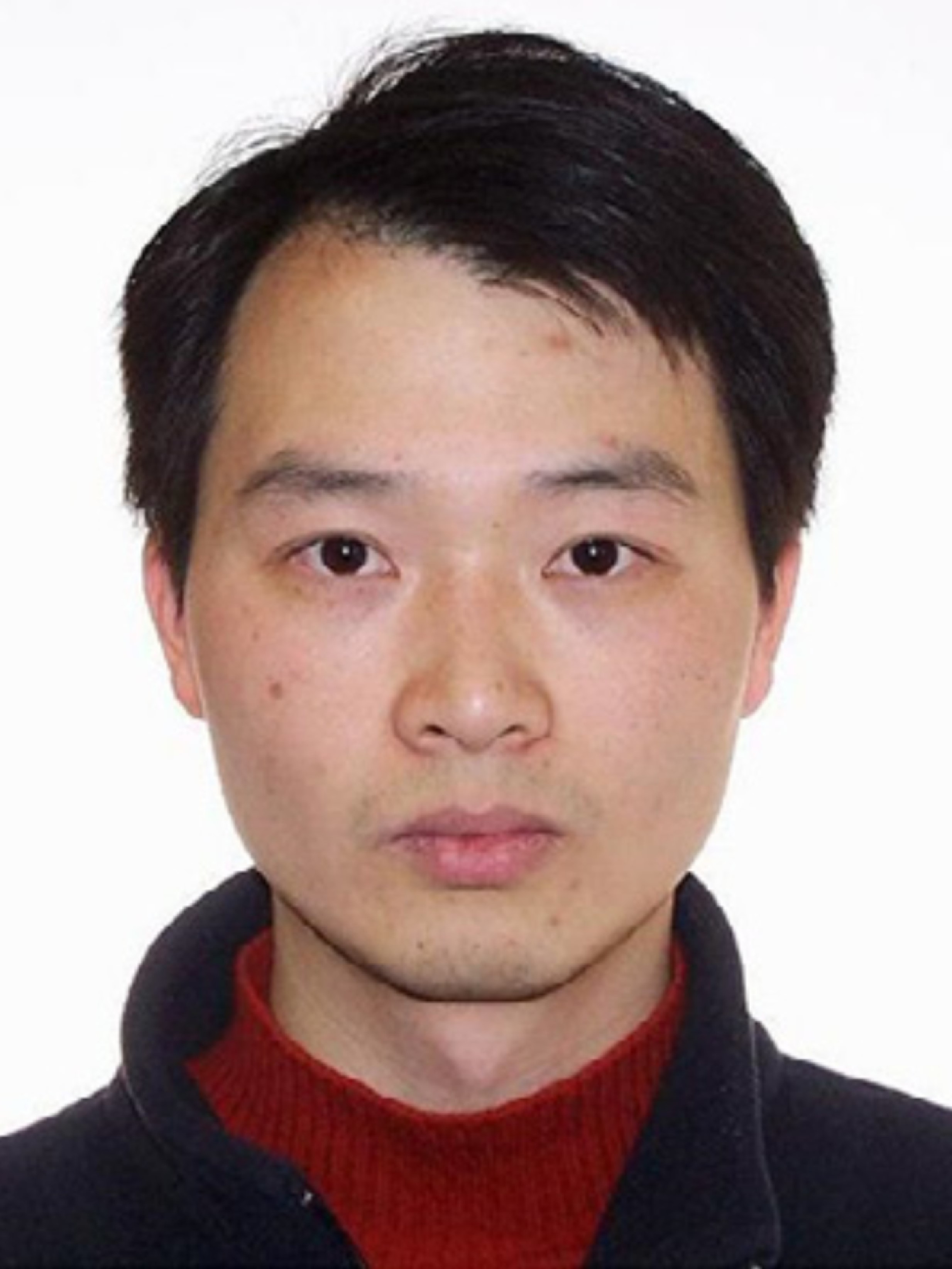}}]{Wei Ni}  (M09-SM15) received the B.E. and Ph.D. degrees in electronic engineering from Fudan University, Shanghai, China, in 2000 and 2005, respectively. He is currently a Team Leader with CSIRO, Sydney, Australia, and an Adjunct Professor with the University of Technology Sydney. He was a Post-Doctoral Research Fellow with Shanghai Jiaotong University from 2005 to 2008, the Deputy Project Manager of the Bell Labs R\&I Center, Alcatel/Alcatel-Lucent from 2005 to 2008, and a Senior Researcher with
Devices Research and Development, Nokia from 2008 to 2009. His research interests include stochastic optimization, game theory, graph theory, and their applications to network and security.
\end{IEEEbiography}

\end{document}